\documentclass[10pt,twocolumn,twoside]{IEEEtran}

\usepackage{graphicx,verbatim,amsmath,amssymb,theorem,array,tabularx}

\newtheorem{lem}{Lemma}

\newtheorem{cor}{Corollary}

\begin{document}


\title{Optimal Power Allocation for Distributed Detection over MIMO Channels in Wireless Sensor Networks}

\author{Xin Zhang, H. Vincent Poor, and Mung Chiang
\thanks{Xin Zhang is with the United Technologies Research Center, East
Hartford, CT 06108. Email: zhangx@utrc.utc.com. H. Vincent Poor
and Mung Chiang are with the Department of Electrical Engineering,
Princeton University, Princeton, NJ 08540. Emails:
poor@princeton.edu; chiangm@princeton.edu.}
\thanks{This paper has been presented in part at the 44th Annual Allerton Conference on Communication,
Control and Computing, Monticello, IL, and at the IEEE Military
Communications Conference 2006, Washington, D.C. }
    \thanks{
    This research was supported in part by the National Science Foundation under
    Grants ANI-03-38807 and CNS-02-25637.}
}

\maketitle

\newcommand{\eqdef}{\stackrel{\Delta}{=}}


\begin{abstract}
In distributed detection systems with wireless sensor networks,
the communication between sensors and a fusion center is not
perfect due to interference and limited transmitter power at the
sensors to combat noise at the fusion center's receiver. The
problem of optimizing detection performance with such imperfect
communication brings a new challenge to distributed detection. In
this paper, sensors are assumed to have independent but
nonidentically distributed observations, and a
multi-input/multi-output (MIMO) channel model is included to
account for imperfect communication between the sensors and the
fusion center. The J-divergence between the distributions of the
detection statistic under different hypotheses is used as a
performance criterion in order to provide a tractable analysis.
Optimizing the performance (in terms of the J-divergence) with
individual and total transmitter power constraints on the sensors
is studied, and the corresponding power allocation scheme is
provided. It is interesting to see that the proposed power
allocation is a tradeoff between two factors, the communication
channel quality and the local decision quality. For the case with
orthogonal channels under certain conditions, the power allocation
can be solved by a weighted water-filling algorithm. Simulations
show that, to achieve the same performance, the proposed power
allocation in certain cases only consumes as little as 25 percent
of the total power used by an equal power allocation scheme.
\end{abstract}


\section{Introduction}
\label{sec:intro}  

Wireless sensor networks (WSN) have received considerable
attention recently. Event monitoring is a typical application of
wireless sensor networks. In event monitoring, a number of sensors
are deployed over a region where some phenomenon is to be
monitored. Each sensor collects and possibly processes data about
the phenomenon and transmits its observation or a summary of its
observation to a fusion center (FC). The FC makes a global
decision about the state of the phenomenon based on the received
data from the sensors, and possibly triggers an appropriate
action.

The essential part of event monitoring is a detection problem,
i.e., the FC needs to detect the state of the phenomenon under
observation. In wireless sensor networks, due to power and
communication constraints, sensors are often required to process
their observations and transmit only summaries of their own
findings to an FC. In this case, the detection problem associated
with event monitoring becomes distributed detection (also called
decentralized detection).

Distributed detection is obviously suboptimal relative to its
centralized counterpart. However, energy, communication bandwidth,
and reliability may favor the use of distributed detection
systems. Distributed detection has been studied for several
decades. Particularly, the design of optimal and suboptimal local
decision and fusion rules has been extensively investigated.
Tsitsiklis \cite{Tsitsiklis 1993}, Varshney \cite{Varshney_book},
Viswanathan and Varshney \cite{Viswanathan Varshney 1997}, and
Blum {\em et al.} \cite{Blum Kassam Poor 1997} provide excellent
reviews of the early work as well as extensive references.

However, most of these studies assume that a finite valued summary
of a sensor is perfectly transmitted to an FC, i.e., no error
occurs during the transmission. In distributed detection systems
based on wireless sensor networks, this assumption may fail due to
interference and limited transmitter power at sensors to combat
receiver noise at the FC. The problem of optimizing detection
performance with imperfect communications between the sensors and
the FC over wireless channels brings a new challenge to
distributed detection.

Rago {\em et al.} \cite{Rago Willett Bar-Shalom 1996} consider a
``censoring'' or ``send/no-send'' idea. The sensors may choose to
transmit data or keep silent according to a total communication
rate constraint and values of their local likelihood ratios.
Predd, Kulkarni and Poor \cite{Predd Kulkarni Poor 2006} examine a
related protocol for the problem of distributed learning. Duman
and Salehi \cite{Duman Salehi 1998} introduce a multiple access
channel model to account for noise and interference in data
transmission, and optimal quantization points (in the
person-by-person sense) were obtained on the original observations
through a numerical procedure. Chen and Willett \cite{Chen Willett
T-IT 2005} assume a general orthogonal channel model from the
local sensors to the FC and investigate the optimality of the
likelihood ratio test (LRT) for local sensor decisions. Chen {\em
et al.} \cite{Chen Jiang Kasetkasem Varshney T-SP 2004} formulate
the parallel fusion problem with a fading channel with
instantaneous channel state information (CSI) and derive the
optimal likelihood ratio (LR)-based fusion rule with binary local
decisions. Niu {\em et al.} \cite{Niu Chen Varshney T-SP 2006}
extend the results of \cite{Chen Jiang Kasetkasem Varshney T-SP
2004} to the case without instantaneous CSI. Note that both
\cite{Chen Jiang Kasetkasem Varshney T-SP 2004} and \cite{Niu Chen
Varshney T-SP 2006} assume orthogonal channels between the sensors
and the FC.

Chamberland and Veeravalli \cite{Chamberland Veeravalli
2003}\cite{Chamberland Veeravalli 2004}\cite{Chamberland
Veeravalli 2006} provide asymptotic results for distributed
detection in power (or equivalently, capacity) constrained
wireless sensor networks. More specifically, \cite{Chamberland
Veeravalli 2003} shows that, when the sensors have i.i.d. Gaussian
or exponential observations and the sensors and the FC are
connected with a multiple access channel with capacity $R$, having
identical binary local decision rules at the sensors is optimal in
the asymptotic regime where the observation interval goes to
infinity. \cite{Chamberland Veeravalli 2004} considers a similar
problem but with a total power constraint instead of a channel
capacity constraint, and shows that using identical local decision
rules at the sensors is optimal for i.i.d. observations.
\cite{Chamberland Veeravalli 2006} considers the detection of
1-dimensional spatial Gaussian stochastic processes. An
amplify-and-relay communication strategy with power constraint is
used and the channels are orthogonal with equal received signal
power from each sensor. They assume sensors are scattered along
1-dimensional space and have correlated observations of the
Gaussian stochastic processes. The tradeoff between sensor density
and the quality of information provided by each sensor is studied
using an asymptotic analysis.

Liu and Sayeed \cite{Liu Sayeed T-SP 2007} and Mergen {\em et al.}
\cite{Mergen Naware Tong T-SP 2007} propose the use of type based
multiple access (TBMA) to transmit local information from the
sensors to the FC, and present a performance analysis of detection
at the FC. The results of \cite{Liu Sayeed T-SP 2007} and
\cite{Mergen Naware Tong T-SP 2007} focus mainly on the case with
i.i.d. observations at the sensors. Jayaweera \cite{Jayaweera T-SP
2007} studies the fusion performance of distributed stochastic
Gaussian signal detection with i.i.d. sensor observations,
assuming an amplify-and-relay scheme.

Chamberland and Veeravalli \cite{Chamberland Veeravalli 2007}
provide a survey of much of the recent progress in distributed
detection in wireless sensor networks with resource constraints.

In this paper, we propose a distributed detection system
infrastructure with a virtual multi-input multi-output (MIMO)
channel to account for non-ideal communications between a finite
number of sensors and an FC. Our analysis does not consider an
infinite number of sensors because in many practical cases, only a
few tens of sensors are used. We assume the sensors have {\em
independent but nonidentically distributed} observations, so they
have different local decision qualities. Each sensor has an
individual transmitting power constraint, and there is also a
joint power constraint on the total amount of power that the
sensors can expend to transmit their local decisions to the FC.
The goal is to optimally distribute the joint power budget among
the sensors so that the detection performance at the FC is
optimized.

The J-divergence between the distributions of the detection
statistic under different hypotheses is used as a performance
index instead of the probability of error in order to provide a
more tractable analysis. A power allocation scheme is developed
with respect to the J-divergence criterion, and in-depth analysis
of the special case of orthogonal channels is provided. The
proposed power allocation is shown to be a tradeoff between two
factors, the quality of the communication channel and the quality
of the local decisions of the sensors. As will be shown in the
simulations, to achieve the same performance in certain cases, the
power allocation developed in this paper consumes as little as 25
percent of the total power used by an equal power allocation
scheme.

This paper differentiates from previous work in the following
aspects.
\begin{itemize}
\item A system with the sensors and the FC connected by a virtual
MIMO channel is considered. \item The sensors have independent but
nonidentically distributed observations, and hence they have
different local decision qualities. \item We develop the power
allocation scheme for a finite number of sensors rather than
asymptotically. \item To improve global detection performance
within a power budget, we focus on how to efficiently and
effectively transmit the local sensor decisions to the FC rather
than how to design local and global decision rules. \item The
proposed power allocation includes tradeoffs between communication
channel quality and local decision quality.
\end{itemize}

The rest of the paper is organized as follows. In Section
\ref{sec:models}, we introduce a distributed detection system
infrastructure with a MIMO channel model. In Section
\ref{sec:powerallocations}, we develop the optimal power
allocation scheme with respect to the J-divergence performance
index. In Section \ref{sec: power allocation special case}, we
study a special case in which the sensors transmit data to the FC
over orthogonal channels. In Section \ref{sec:simulations}, we
provide numerical examples to illustrate the proposed power
allocation. We conclude the paper in Section
\ref{sec:conclusions}.

\section{Models}
\label{sec:models} 
\begin{figure}[t]
\begin{center}
\includegraphics*[width=0.45\textwidth]{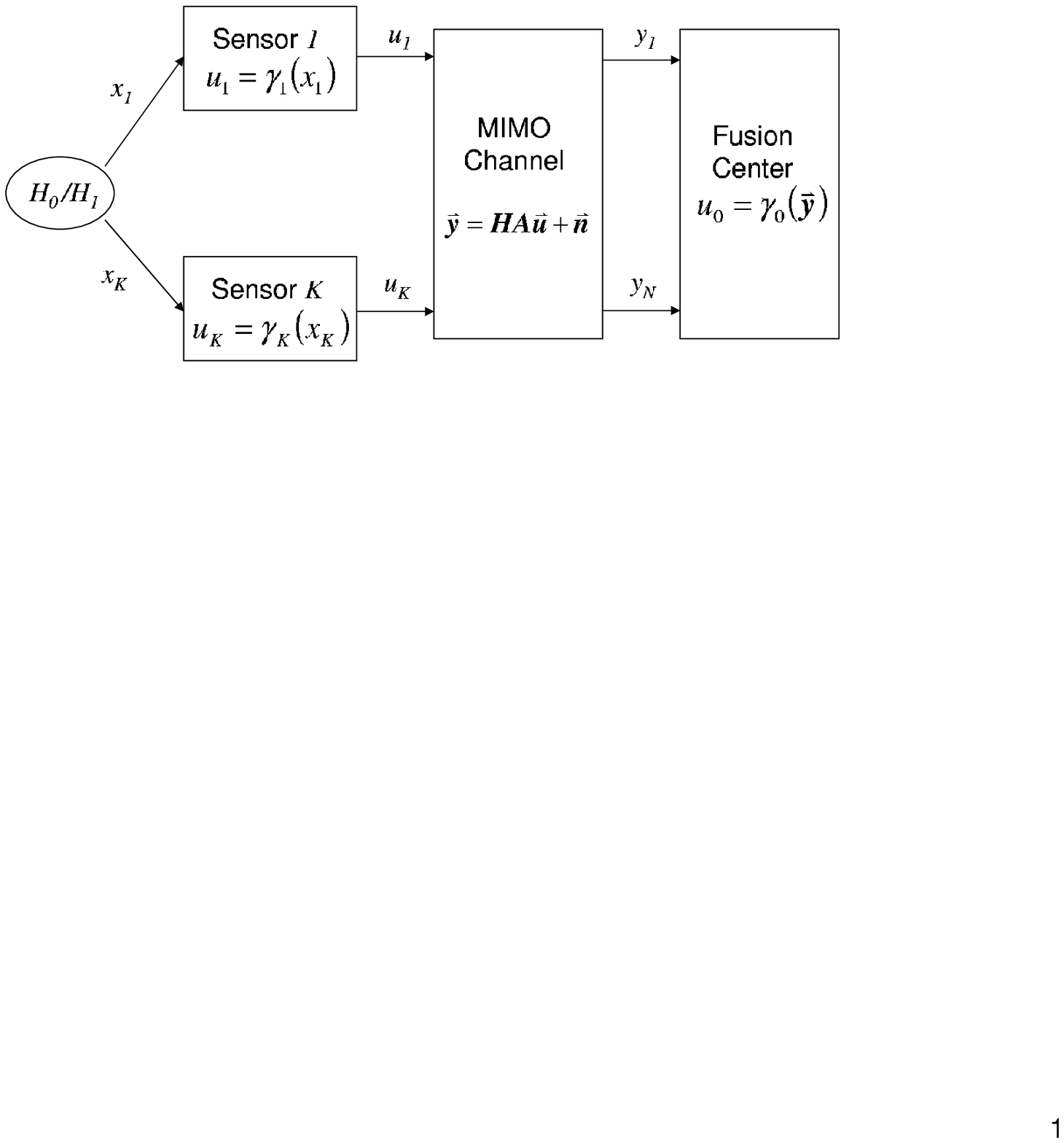}
\end{center}
\caption{Distributed detection system diagram.} \label{fig:system}
\end{figure}

Let us consider a hypothesis testing problem
with two hypotheses $H_0$ and $H_1$, as shown in Figure
\ref{fig:system}. There are $K$ wireless sensors with observations
${\bf x}=[x_1,\cdots,x_K]^T$. The observations are independent of
each other but are not necessarily identically distributed. The
conditional probability density functions of these observations
(conditioned on the underlying hypotheses) are given by $p({\bf
x}|H_i)$ for $i\in\{0,1\}$. The sensors then make local decisions
${\bf u}=[u_1,\cdots,u_K]^T$ according to their local decision
rules:
\begin{equation}
u_k=\gamma_k(x_k)=\begin{cases} 0 & \text{decide $H_0$} \\ 1 &
\text{decide $H_1$} \end{cases},
\end{equation}
where $k=1,\cdots,K$. In this paper, we assume the local sensors
do not communicate with each other, i.e., sensor $k$ makes a
decision independently based only on its own observation $x_k$.
The local decision rules $\gamma_k(\cdot)$ do not have to be
identical, and the false alarm probability and detection
probability of sensor $k$ are given by
\begin{align}
P_{F}(k)&=p(u_k=1|H_0),
\\[-3mm]
\intertext{and}\notag \\[-8mm]
P_{D}(k)&=p(u_k=1|H_1).
\end{align}
We assume the sensors have knowledge of their observation quality
in terms of $P_D$ and $P_F$, which can be obtained by various
standard methods from detection theory \cite{Poor_book}. The joint
conditional density functions of the local decisions are
\begin{align}
\label{eqn:p(u|H0)}p({\bf u}|H_0)&=\prod_{k=1}^K
P_F(k)^{u_k}(1-P_F(k))^{(1-u_k)}, \\[-3mm]
\intertext{and}\notag \\[-9mm]
\label{eqn:p(u|H1)}p({\bf u}|H_1)&=\prod_{k=1}^K
P_D(k)^{u_k}(1-P_D(k))^{(1-u_k)}.
\end{align}
The local decisions are transmitted to an FC through a MIMO
channel, modelled by the following sampled baseband signal model
(see, e.g., \cite{Wang_Poor_book}):
\begin{equation}
{\bf y}={\bf HAu}+{\bf n}, \label{eqn:channel model}
\end{equation}
where ${\bf y}=[y_1,\cdots,y_N]^T$ contains the received signals
at the FC. ${\bf A}={\rm diag}\{a_1,a_2,\cdots,a_K\}$, is a
diagonal matrix, the diagonal elements of which are the amplitudes
of the signals transmitted from the sensors. ${\bf H}$ is the
channel matrix, which is assumed to be deterministic in this
paper\footnote{We focus on the case in which the sensors and the
FC have minimal movement and the environment changes slowly. In
this case, the coherence time \cite{Tse_Viswanath_book} of the
wireless channel can be much longer than the time interval between
two consecutive decisions made by the FC, and instantaneous CSI
can be obtained.}. ${\bf n}$ is an additive noise vector which is
assumed to be Gaussian with zero mean and covariance matrix ${\bf
R}$. We assume that the channel quality (in terms of the
$\mathbf{H}$ and $\mathbf{R}$ matrices) is known at the FC. This
information can be obtained by channel estimation techniques. The
dimension of ${\bf y}$, determined by the receiver design, is $N$,
which does not have to be the same as the number of sensors $K$.
Many different wireless channels and multiple access schemes can
be expressed with the MIMO model in \eqref{eqn:channel model},
including CDMA, TDMA, FDMA, as well as TBMA \cite{Liu Sayeed T-SP
2007}\cite{Mergen Naware Tong T-SP
2007}\cite{Verdu_book}\cite{Tse_Viswanath_book}.

The conditional density function of the received signals ${\bf y}$
at the FC given the transmitted signals ${\bf u}$ from the sensors
is
\begin{align}
\!p({\bf y}|{\bf u})\!&=\!\frac{1}{|2\pi{\bf
R}|^{\frac{1}{2}}}{\rm exp}\!\left[\!-\frac{1}{2} ({\bf
y}\!-\!{\bf HAu})^T{\bf R}^{-1}({\bf y}\!-\!{\bf HAu}) \right].
\label{eqn:p(y|u)}
\end{align}
The conditional density functions of the received signals given
the two hypotheses are
\begin{equation}
p({\bf y}|H_i)=\sum_{\bf u}p({\bf y}|{\bf u})p({\bf u}|H_i),
\label{eqn:cond pdf y}
\end{equation}
where the summation is over all possible values of ${\bf u}$.
The FC applies its fusion rule $\gamma_0(\cdot)$ to ${\bf y}$ to
get a global decision
\begin{equation}
u_0=\gamma_0({\bf y}).
\end{equation}
The system is summarized in Figure \ref{fig:system}. We notice the
Markov property of the system: $\{H_i\} \rightarrow {\bf x}
\rightarrow {\bf u} \rightarrow {\bf y} \rightarrow u_0$ forms a
Markov chain, which is used to derive \eqref{eqn:cond pdf y} and
will be used in the next section.

In this paper, we do not focus on the design of local and global
decision rules to optimize the detection performance at the FC.
Instead, we focus on how to intelligently distribute a total
transmitter power budget $P_{\rm tot}$ among the sensors, by
choosing an amplitude matrix $\mathbf{A}$ within the constraint
${\rm Tr}\left[{\bf AA}^T\right] \le P_{\rm tot}$. There are also
individual power constraint for each sensor, ${\bf A} \preceq
\sqrt{{\bf P}_{\rm max}}$, to account for the maximum output power
at each sensor. Here, $\sqrt{{\bf P}_{\rm max}}$ denotes the
component-wise square root of ${\bf P}_{\rm max}={\rm
diag}\{P_{\rm max}(1),\cdots,P_{\rm max}(K)\}$, where $P_{\rm
max}(k)$ is the transmitting power limit of sensor $k$. The matrix
inequality $\preceq $ means $\sqrt{{\bf P}_{\rm max}}-{\bf A}$ is
positive semidefinite.

\section{Optimal Power Allocation}
\label{sec:powerallocations} In this section, an optimal power
allocation among the sensors in the distributed detection system
described in Section \ref{sec:models} is studied. We first choose
a detection performance metric for our analysis.

There are three categories of commonly used detection performance
metrics \cite{Poor_book}: exact closed-form expressions of the
miss probability $P_M$ (which equals $1-P_D$) and false alarm
probability $P_F$ (or the average error probability $P_e$, if
prior probabilities of the hypotheses are known), distance related
bounds, and asymptotic relative efficiency (ARE).

The natural performance metric, closed-form expressions of $P_M$
and $P_F$ (or $P_e$), is hard to obtain even for centralized
detection. ARE is useful for detection systems under
large-sample-size (long observation duration) and weak signal
conditions. Distance related bounds are upper or lower bounds on
$P_M$ and $P_F$ (or $P_e$), such as the Chernoff bound, the
Bhattacharyya bound, and the J-divergence \cite{Poor_book}.

In this paper, we use distance related bounds, more specifically
the J-divergence, as the performance metric. The J-divergence,
first proposed by Jeffreys \cite{Jeffreys 1946}, is a widely used
metric for detection performance \cite{Kailath
1967}\cite{Kobayashi Thomas 1967}\cite{Kobayashi 1970}\cite{Poor
Thomas 1977}. It provides a lower bound on the detection error
probability $P_e$ \cite{Kobayashi Thomas 1967} via the inequality
\begin{align} \label{eqn:J-divergence
as a lower bound} P_e
> p(H_0)p(H_1)e^{-J/2}.
\end{align}

We choose the J-divergence as the performance metric because it
provides more tractable results in our study, it is closely
related to results in information theory, such as the data
processing lemma \cite{Csiszar_book}, and it is also closely
related to other types of performance metrics. \cite{Kobayashi
1970} shows that the ratio of the J-divergences of two test
statistics is equivalent to the ARE under some circumstances. The
J-divergence and the Bhattacharyya bound both belong to a more
general class of distance measures, the Ali-Silvey class of
distance measures \cite{Ali Silvey 1966}. The J-divergence is the
symmetric version of the Kullback-Leibler (KL) distance
\cite{Cover_book}\cite{Csiszar_book}, and the KL distance is
asymptotically the error exponent of the Chernoff bound from
Stein's lemma \cite{Cover_book}.

The J-divergence between two densities, $p_1$ and $p_0$, is
defined as
\begin{align}
\label{eqn:def J-divergence} J(p_1,p_0)=D(p_1||p_0)+D(p_0||p_1),
\end{align}
where $D(p_1||p_0)$ is the (non-symmetric) KL distance between
$p_1$ and $p_0$. $D(p_1||p_0)$ and $D(p_0||p_1)$ are defined as
\begin{align}
D(p_i||p_j)=\int {\rm log}\left(\frac{p_i}{p_j}\right) p_i.
\label{eqn:def KL distance}
\end{align}

There is a well-known data processing lemma on the KL distance
along a Markov chain \cite[LEMMA 3.11]{Csiszar_book}:
\begin{lem}
\label{thm: data processing} The KL distance is non-increasing
along the Markov chain ${\bf x} \rightarrow {\bf u} \rightarrow
{\bf y}$, i.e.,
\begin{align}
D(p({\bf x}|H_1)||p({\bf x}|H_0)) &\ge D(p({\bf u}|H_1)||p({\bf
u}|H_0)), \\[-3mm]
\intertext{and}\notag \\[-8mm] D(p({\bf u}|H_1)||p({\bf
u}|H_0)) &\ge D(p({\bf y}|H_1)||p({\bf y}|H_0)).
\end{align}
\end{lem}
This result can be easily generalized to the J-divergence with the
following corollary.
\begin{cor}
\label{cor: data processing J-divergence} The J-divergence is
non-increasing along the Markov chain ${\bf x} \rightarrow {\bf u}
\rightarrow {\bf y}$, i.e.,
\begin{align}
J(p({\bf x}|H_1),p({\bf x}|H_0)) &\ge J(p({\bf u}|H_1),p({\bf
u}|H_0)), \\[-3mm]
\intertext{and}\notag \\[-8mm] J(p({\bf u}|H_1),p({\bf
u}|H_0)) &\ge J(p({\bf y}|H_1),p({\bf y}|H_0)).
\end{align}\end{cor}
Corollary \ref{cor: data processing J-divergence} tells us that a
performance upper bound of the detection at the FC is provided by
$J(p({\bf u}|H_1),p({\bf u}|H_0))$. This can be achieved only when
there are perfect data transmissions from the sensors to the FC,
i.e., the FC receives ${\bf u}$ with no error.

Recall that our goal is to optimize the detection performance at
the FC. This now translates into maximization of the J-divergence
between the two densities of the received signals ${\bf y}$, with
respect to the underlying hypotheses. The optimal power allocation
is thus the solution to the following optimization problem:
\begin{align}
\label{eqn:J optimization}
\max_{{\bf A}} & \quad J(p({\bf y}|H_1),p({\bf y}|H_0)), \\
{\rm s.t.} & \quad {\rm Tr}\left[{\bf AA}^T\right] \le P_{\rm
tot}, \nonumber
\\ & \quad {\bf 0} \preceq {\bf
A} \preceq \sqrt{{\bf P}_{\rm max}}, \nonumber
\end{align}
where the J-divergence $J(p({\bf y}|H_1),p({\bf y}|H_0))$ is given
by
\begin{align}
\label{eqn:J expression} &J(p({\bf y}|H_1),p({\bf y}|H_0))
\nonumber \\
&= \int_{\bf y}d{\bf y}\left[p({\bf y}|H_1)-p({\bf y}|H_0)\right]
{\rm log}\frac{p({\bf y}|H_1)}{p({\bf y}|H_0)}.
\end{align}
The density functions $p({\bf y}|H_i)$, $i\in\{0,1\}$, are given
by \eqref{eqn:p(y|u)}--\eqref{eqn:cond pdf y}.

It can be seen that the conditional density functions $p({\bf
y}|H_i)$ are Gaussian mixtures. Unfortunately, the J-divergence
between two Gaussian mixture densities does not have a general
closed-form expression \cite{Moreno GM KL 2003}\cite{Singer
Warmuth GM KL 1998}. In order to present the objective function in
\eqref{eqn:J optimization} in closed-form, approximations must be
made. An upper bound has been suggested in \cite{Singer Warmuth GM
KL 1998} based on the log-sum inequality \cite{Cover_book}.
However, this upper bound is not suitable for the study here,
since the dependence on the power of transmitted signals is lost
in the bound.

In this paper, the J-divergence of two Gaussian mixture densities
$p({\bf y}|H_i)$ is approximated by the J-divergence of two
Gaussian densities $p_{g}({\bf y}|H_i)={\cal N}({\bf y};{\bf
\mu}_i,{\bf \Sigma}_i)$. The parameters of the Gaussian densities
are provided by moment matching, i.e.,
\begin{align}
\label{eqn:G approx mean}{\bf \mu}_i&=\int_{\bf y}{\bf y}p({\bf
y}|H_i)d{\bf y}, \\[-3mm]
\intertext{and}\notag \\[-8mm] {\bf
\Sigma}_i&=\int_{\bf y}[{\bf y}-{\bf \mu}_i][{\bf y}-{\bf
\mu}_i]^Tp({\bf y}|H_i)d{\bf y}, \label{eqn:G approx var}
\end{align}
for $i\in\{0,1\}$. That is, a Gaussian mixture density is
approximated by a Gaussian density with the same mean and variance
as the Gaussian mixture density.

Obviously the quality of this approximation will directly affect
the analysis in this paper and the difference between the optimal
scheme and the proposed scheme, which is optimal for the
approximated cases. It can be seen from
\eqref{eqn:p(y|u)}--\eqref{eqn:cond pdf y} that when ${\rm
Tr}[({\bf HA})^T{\bf R}^{-1}({\bf HA})] \rightarrow 0$, the
Gaussian mixture density in \eqref{eqn:cond pdf y} approaches a
Gaussian distribution. So, we can predict the approximation will
work well for the low SNR cases, and simulations in Section
\ref{sec:simulations} show that it still works well even with
received SNR as high as 10-12 dB.

We next calculate the means and covariance matrices of the
Gaussian densities $p_{g}({\bf y}|H_i)$, $i\in\{0,1\}$. From
\eqref{eqn:cond pdf y}, \eqref{eqn:G approx mean}, and the Markov
property of the system,
\begin{align}
{\bf \mu}_i&=\int_{\bf y}{\bf y}p({\bf y}|H_i)d{\bf y}, \nonumber \\
&=\int_{\bf y}{\bf y}\sum_{\bf u}p({\bf y}|{\bf u})p({\bf
u}|H_i)d{\bf y}, \nonumber \\
&=\sum_{\bf u}p({\bf u}|H_i)\int_{\bf y}{\bf y}p({\bf y}|{\bf
u})d{\bf y}.
\end{align}
Recall that $p({\bf y}|{\bf u})$ is a Gaussian density with mean
${\bf HAu}$, as shown in \eqref{eqn:p(y|u)}, so
\begin{align}
{\bf \mu}_i&=\sum_{\bf u}{\bf HAu}p({\bf u}|H_i).
\end{align}
By applying \eqref{eqn:p(u|H0)} and \eqref{eqn:p(u|H1)}, we have
\begin{align}
\label{eqn:mean of single Gaussian} {\bf \mu}_i &= {\bf HA
\beta}_i,
\end{align}
for $i\in\{0,1\}$,
\begin{align}
{\bf \beta}_1 &= \sum_{\bf u}{\bf u}p({\bf u}|H_1) =
\left[P_D(1),\cdots,P_D(K) \right]^T, \\[-3mm]
\intertext{and}\notag \\[-8mm] {\bf \beta}_0 &= \sum_{\bf u}{\bf u}p({\bf u}|H_0) =
\left[P_F(1),\cdots,P_F(K) \right]^T.
\end{align}
Similarly, from \eqref{eqn:G approx var} and the Markov property
of the system, we have
\begin{align}
{\bf \Sigma}_i&=\int_{\bf y}[{\bf y}-{\bf \mu}_i][{\bf y}-{\bf
\mu}_i]^Tp({\bf y}|H_i)d{\bf y}, \nonumber \\
&=\int_{\bf y}[{\bf y}-{\bf \mu}_i][{\bf y}-{\bf
\mu}_i]^T\sum_{\bf u}p({\bf y}|{\bf u})p({\bf
u}|H_i)d{\bf y}, \nonumber \\
&=\sum_{\bf u}p({\bf u}|H_i)\int_{\bf y}[{\bf y}-{\bf HAu}+{\bf
HAu}-{\bf \mu}_i] \nonumber \\
& \qquad \qquad \cdot [{\bf y}-{\bf HAu}+{\bf HAu}-{\bf \mu}_i]^Tp({\bf y}|{\bf u})d{\bf y}, \nonumber \\
&={\bf R}+\sum_{\bf u}p({\bf u}|H_i)[{\bf HAu}-{\bf \mu}_i][{\bf
HAu}-{\bf \mu}_i]^T.
\end{align}
The last step follows because $p({\bf y}|{\bf u})$ is a Gaussian
density with mean ${\bf HAu}$ and covariance matrix ${\bf R}$.
Applying \eqref{eqn:p(u|H0)} and \eqref{eqn:p(u|H1)}, and after
some algebra, we obtain
\begin{align}
\label{eqn:covariance of single Gaussian} {\bf \Sigma}_i &= {\bf
R} + {\bf HAB}_i{\bf A}^T{\bf H}^T,
\end{align}
where
\begin{align}
\label{eqn:def B_1}{\bf B}_1 &= {\rm diag} \left\{P_D(1)[1-P_D(1)]
,\cdots,P_D(K)[1-P_D(K)] \right\}, \\[-3mm]
\intertext{and}\notag \\[-6mm]
\label{eqn:def B_0}{\bf B}_0 &= {\rm diag} \left\{P_F(1)[1-P_F(1)]
,\cdots,P_F(K)[1-P_F(K)] \right\}.
\end{align}

We next derive the J-divergence between the Gaussian densities,
$J(p_g({\bf y}|H_1),p_g({\bf y}|H_0))$. From the definition of the
J-divergence and the KL distance in \eqref{eqn:def
J-divergence}--\eqref{eqn:def KL distance}, we have
\begin{align}
&J(p_g({\bf y}|H_1),p_g({\bf y}|H_0)) \nonumber \\ &= \int_{\bf
y}d{\bf y}\left[p_g({\bf y}|H_1)-p_g({\bf y}|H_0)\right] {\rm
log}\frac{p_g({\bf y}|H_1)}{p_g({\bf y}|H_0)}.
\end{align}
Using the fact that $p_g({\bf y}|H_i)$ are Gaussian densities
${\cal N}({\bf y};{\bf \mu}_i,{\bf \Sigma}_i)$, $i\in\{0,1\}$,
after some algebra, we obtain
\begin{align}
& J(p_g({\bf y}|H_1),p_g({\bf y}|H_0)) \nonumber \\
&= \frac{1}{2}{\rm
Tr}\Big[\Sigma_0\Sigma_1^{-1}+\Sigma_1\Sigma_0^{-1}\nonumber \\
& \qquad + \left(\Sigma_1^{-1}+\Sigma_0^{-1}\right)
\left(\mu_1-\mu_0\right) \left(\mu_1-\mu_0\right)^T \Big]-N,
\end{align}
where $N$ is the dimension of the received signal vector ${\bf y}$
at the FC. Using the means ${\bf \mu}_i$ and covariance matrices
${\bf \Sigma}_i$ in \eqref{eqn:mean of single Gaussian} and
\eqref{eqn:covariance of single Gaussian}, and after some algebra,
we have,
\begin{align}
\label{eqn:J approx expression} & J(p_g({\bf y}|H_1),p_g({\bf
y}|H_0)) \nonumber \\
& = \frac{1}{2}{\rm Tr}\Big[\left[{\bf R} + {\bf HA}({\bf
B}_0+{\bf \beta \beta}^T){\bf A}^T{\bf H}^T
\right] \nonumber \\
& \qquad \qquad \cdot \left[{\bf R} + {\bf HAB}_1{\bf A}^T{\bf H}^T \right]^{-1} \Big] \nonumber \\
& \qquad + \frac{1}{2}{\rm Tr}\Big[ \left[{\bf R} + {\bf HA}({\bf
B}_1+{\bf \beta \beta}^T){\bf A}^T{\bf H}^T \right] \nonumber \\
& \qquad \qquad \cdot \left[{\bf R} + {\bf HAB}_0{\bf A}^T{\bf
H}^T \right]^{-1} \Big]-N,
\end{align}
where ${\bf \beta}={\bf \beta}_1-{\bf \beta}_0$.

The approximated optimal power allocation is the solution to the
following optimization problem:
\begin{align}
\label{eqn:J approx optimization}
\max_{{\bf A}} & \quad J(p_g({\bf y}|H_1),p_g({\bf y}|H_0)),
\\
{\rm s.t.} & \quad {\rm Tr}\left[{\bf AA}^T\right] \le P_{\rm
tot}, \nonumber
\\ & \quad {\bf 0} \preceq {\bf
A} \preceq \sqrt{{\bf P}_{\rm max}}. \nonumber
\end{align}
For the objective function given in \eqref{eqn:J approx
expression}, the optimization is over the amplitude matrix ${\bf
A}$, or equivalently the power allocation among the sensors. The
optimization problem can be solved by various constrained
optimization techniques, and in the simulations we use the
interior point method \cite{Bertsekas99book}\cite{Boyd_book}.

\section{Special case with orthogonal channels}
\label{sec: power allocation special case} 
\begin{figure}[tb]
\includegraphics[width=0.45\textwidth]{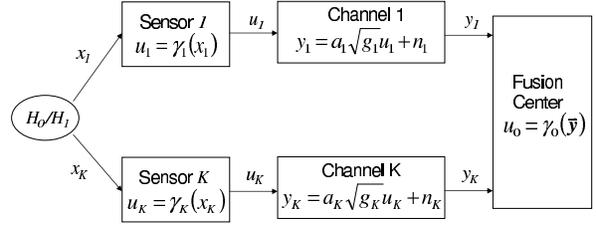}
\caption{\label{fig:system_specialcase} Distributed detection
system with orthogonal channels.}
\end{figure}

A special case of the
distributed detection system depicted in Figure \ref{fig:system},
is that in which all of the sensors have orthogonal channels for
communication with the FC. A system diagram for this case is shown
in Figure \ref{fig:system_specialcase}.

\begin{figure}[tb]
\includegraphics[width=0.38\textwidth]{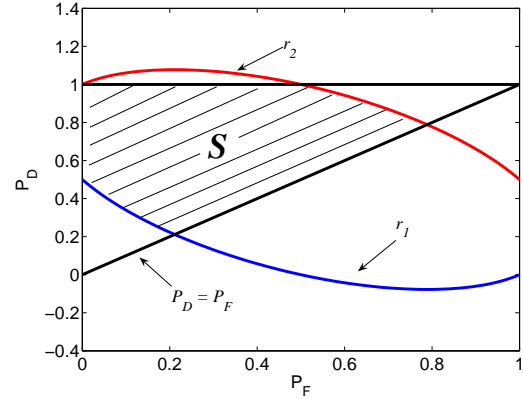}
\caption{\label{fig:PD_PF_S0_S1} Illustration of region
$\mathbf{\cal S}$.}
\end{figure}

Compared to the system in Figure \ref{fig:system}, this special
case has
\begin{align}
\label{eqn:specialcase H R} {\bf H}&={\rm diag} \left\{\sqrt{g_1},\cdots,\sqrt{g_K} \right\}, \\
{\bf R}&=\sigma^2 {\bf I}_K,
\end{align}
where ${\bf I}_K$ is a $K$-by-$K$ identity matrix, and the noises
in all the channels are independent and have the same variance
$\sigma^2$. Here, $g_j$ is the channel power gain for sensor $j$.
By substituting the above two matrices into the optimization
problem in \eqref{eqn:J approx expression}--\eqref{eqn:J approx
optimization}, after some algebra, the power allocation for this
special case reduces to the solution to the following optimization
problem,
\begin{align}
\label{eqn:J approx optimization special case}
\max_{\{P_1,\cdots,P_K\}} & \quad J(P_1,\cdots,P_K) \nonumber \\
& = \sum_{j=1}^K \left[\frac{\sigma^2+\alpha_F(j)g_jP_j}
{\sigma^2+\beta_F(j)g_jP_j}+\frac{\sigma^2+\alpha_D(j)g_jP_j}{\sigma^2+\beta_D(j)g_jP_j}
\right], \\
{\rm s.t.} & \quad \sum_{j=1}^K P_j \le P_{\rm tot}, \nonumber
\\ & \quad 0\le P_j \le P_{\rm max}(j), \qquad j=1,\cdots,K, \nonumber
\end{align}
where
\begin{align}
\label{eqn:alpha_D beta_D}
\alpha_F(j) &= P_F(j)(1-P_D(j))+P_D(j)(P_D(j)-P_F(j)), \\
\alpha_D(j) &= P_D(j)(1-P_F(j))-P_F(j)(P_D(j)-P_F(j)), \\
\beta_F(j) &= P_D(j)(1-P_D(j)), \\[-1mm]
\intertext{and}\notag \\[-7mm] \beta_D(j) &=
P_F(j)(1-P_F(j)). \label{eqn:alpha_D beta_D end}
\end{align}
Note that $P_j=a_j^2$ is the power allocated to sensor $j$ for
transmitting its findings to the FC. The objective function is
fully decoupled, a direct result of the orthogonal channels
between the sensors and the FC.

The first order partial derivative of $J(P_1,\cdots,P_K)$ with
respect to $P_j$ is given by
\begin{align} \label{eqn:1st derivative of J(P)} &\frac{\partial}{\partial
P_j}J(P_1,\cdots,P_K) \nonumber \\
&=\frac{(\alpha_F(j)-\beta_F(j))\sigma^2g_j}
{(\sigma^2+\beta_F(j)g_jP_j)^2}+\frac{(\alpha_D(j)-\beta_D(j))\sigma^2g_j}{(\sigma^2+\beta_D(j)g_jP_j)^2}.
\end{align}
It has a interesting property as stated by the following lemma.
\begin{lem}
\label{lem: special case J non decreasing} The first order
derivative of the objective function $J(P_1,\cdots,P_K)$ with
respect to $P_j$ is always nonnegative at any valid power
allocation point $P_i \ge 0$. That is
\begin{align}
\label{eqn:J approx derivative expression}
&\frac{\partial}{\partial P_j} J(P_1,\cdots,P_K)\Big |_{P_i\ge0}
\ge 0 .
\end{align}
\end{lem}
\begin{proof}
See Appendix \ref{proof:lem 4}.
\end{proof}
Lemma \ref{lem: special case J non decreasing} tells us that the
objective function in \eqref{eqn:J approx optimization special
case} is nondecreasing with increasing power budget $P_{\rm tot}$.
Since we are maximizing a nondecreasing function, the optimal
point is always at the constraint boundary, i.e., $\sum_{j=1}^K
P_j = P_{\rm tot}$, or $P_j = P_{\rm max}(j), \; j=1,\cdots,K$.
This result is intuitively plausible since it makes full use of
the power budget.

Practical sensors should always have $P_D>P_F$, since, if
$P_D=P_F$, the sensors do not provide useful information. With
this condition, we can easily prove the following corollary.

\begin{cor}
\label{cor: positive 1st order derivative} If $P_D(j)>P_F(j)$,
then the first order derivative of the objective function
$J(P_1,\cdots,P_K)$ with respect to $P_j$ is always strictly
positive at any valid power allocation point $P_i\ge0$.
\end{cor} Corollary \ref{cor: positive 1st order derivative} tells
us that there is no stationary point inside the constraint
boundary, so gradient based optimization techniques will not get
stuck.

The second order partial derivative of $J(P_1,\cdots,P_K)$ with
respect to $P_j$ is given by
\begin{align}
\label{eqn:2nd derivative of J(P)} &\frac{\partial^2}{\partial
P_j^2}J(P_1,\cdots,P_K) \nonumber \\
&=-\sigma^2g_j\left[\frac{\left[\alpha_F(j)-\beta_F(j)\right]\beta_F(j)}
{\left[\sigma^2+\beta_F(j)g_jP_j\right]^3}\right. \notag
\\
& \qquad \qquad \qquad
\left.+\frac{\left[\alpha_D(j)-\beta_D(j)\right]\beta_D(j)}
{\left[\sigma^2+\beta_D(j)g_jP_j\right]^3}\right]
\nonumber \\
&=-\sigma^2g_j\left[\frac{C_0\sigma^6+C_1\sigma^4g_jP_j+C_2\sigma^2g_j^2P_j^2+C_3g_j^3P_j^3}
{\left[\sigma^2+\beta_F(j)g_jP_j\right]^3\left[\sigma^2+\beta_D(j)g_jP_j\right]^3}\right],
\end{align}
where,
\begin{align}
\label{eqn:def C0--C3}
C_0&=\beta_F(j)\Big[\alpha_F(j)-\beta_F(j)\Big]+\beta_D(j)\Big[\alpha_D(j)-\beta_D(j)\Big],
\\
C_1&=\beta_F(j)\beta_D(j)\Big[\alpha_F(j)-\beta_F(j)+\alpha_D(j)-\beta_D(j)\Big], \\
C_2&=\beta_F(j)\beta_D(j)\Big[\beta_D(j)\left[\alpha_F(j)-\beta_F(j)\right]\nonumber
\\&\qquad \qquad
+\beta_F(j)\left[\alpha_D(j)-\beta_D(j)\right]\Big], \\[-3mm]
\intertext{and}\notag \\[-8mm]
C_3&=\beta_F(j)\beta_D(j)\Big[\beta_D(j)^2\left[\alpha_F(j)-\beta_F(j)\right]\nonumber
\\&\qquad \qquad
+\beta_F(j)^2\left[\alpha_D(j)-\beta_D(j)\right]\Big].
\end{align}
The second order partial derivative in \eqref{eqn:2nd derivative
of J(P)} is not always nonpositive, which means the objective
function $J(P_1,\cdots,P_K)$ is not always concave. However, the
following lemma specifies the region in terms of local sensor
observation quality, where the second order derivative of the
objective function is indeed nonpositive. We again assume that
practical sensors have $P_D>P_F$.
\begin{lem}
\label{lem: special case J second derivative} The second order
partial derivative of the objective function,
$\frac{\partial^2}{\partial P_j^2}J(P_1,\cdots,P_K)\le 0$, for any
allocated power $P_j\ge0$, if and only if
$\big(P_D(j),P_F(j)\big)\in \mathbf{\cal S}$, where $\mathbf{\cal
S}$ is defined by
\begin{align}\Big\{\mathbf{\cal
S}(P_D,P_F)\Big|&\frac{3}{4}-\frac{1}{2}P_F-\frac{1}{4}\sqrt{1+12P_F-12P_F^2}
\nonumber \\
& \;\le P_D \le
\frac{3}{4}-\frac{1}{2}P_F+\frac{1}{4}\sqrt{1+12P_F-12P_F^2},
\notag \\ & \qquad \qquad 0 \le P_F < P_D\le 1 \Big\}.
\end{align}
\end{lem}
\begin{proof}
See Appendix \ref{proof:lem 5}.
\end{proof}
Region $\mathbf{\cal S}$ is depicted in Figure
\ref{fig:PD_PF_S0_S1}, in which
$r_1=\frac{3}{4}-\frac{1}{2}P_F-\frac{1}{4}\sqrt{1+12P_F-12P_F^2}$
and
$r_2=\frac{3}{4}-\frac{1}{2}P_F+\frac{1}{4}\sqrt{1+12P_F-12P_F^2}$.
We will show that, if all the sensors operate in region
$\mathbf{\cal S}$, the power allocation can be solved by a
weighted waterfilling algorithm.

\begin{figure*}[tb]
\begin{minipage}{0.33\linewidth}
\includegraphics[width=\textwidth]{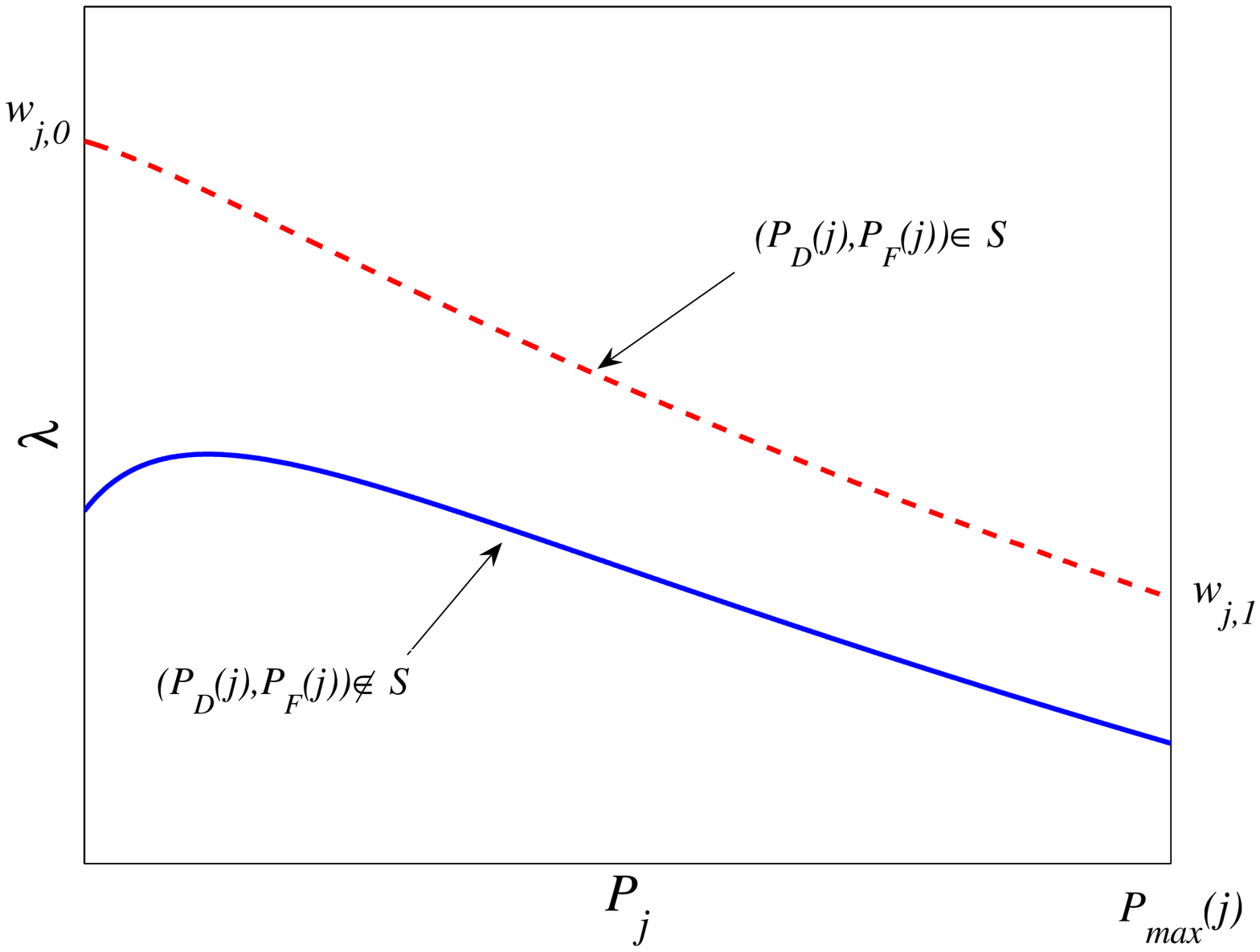}
\caption{\label{fig:lambda_Pj} $\lambda$ as a function of $P_j$
for sensors operating in or not in region $\mathbf{\cal S}$.}
\end{minipage}
\hfill
\begin{minipage}{0.31\linewidth}
\includegraphics[width=\textwidth]{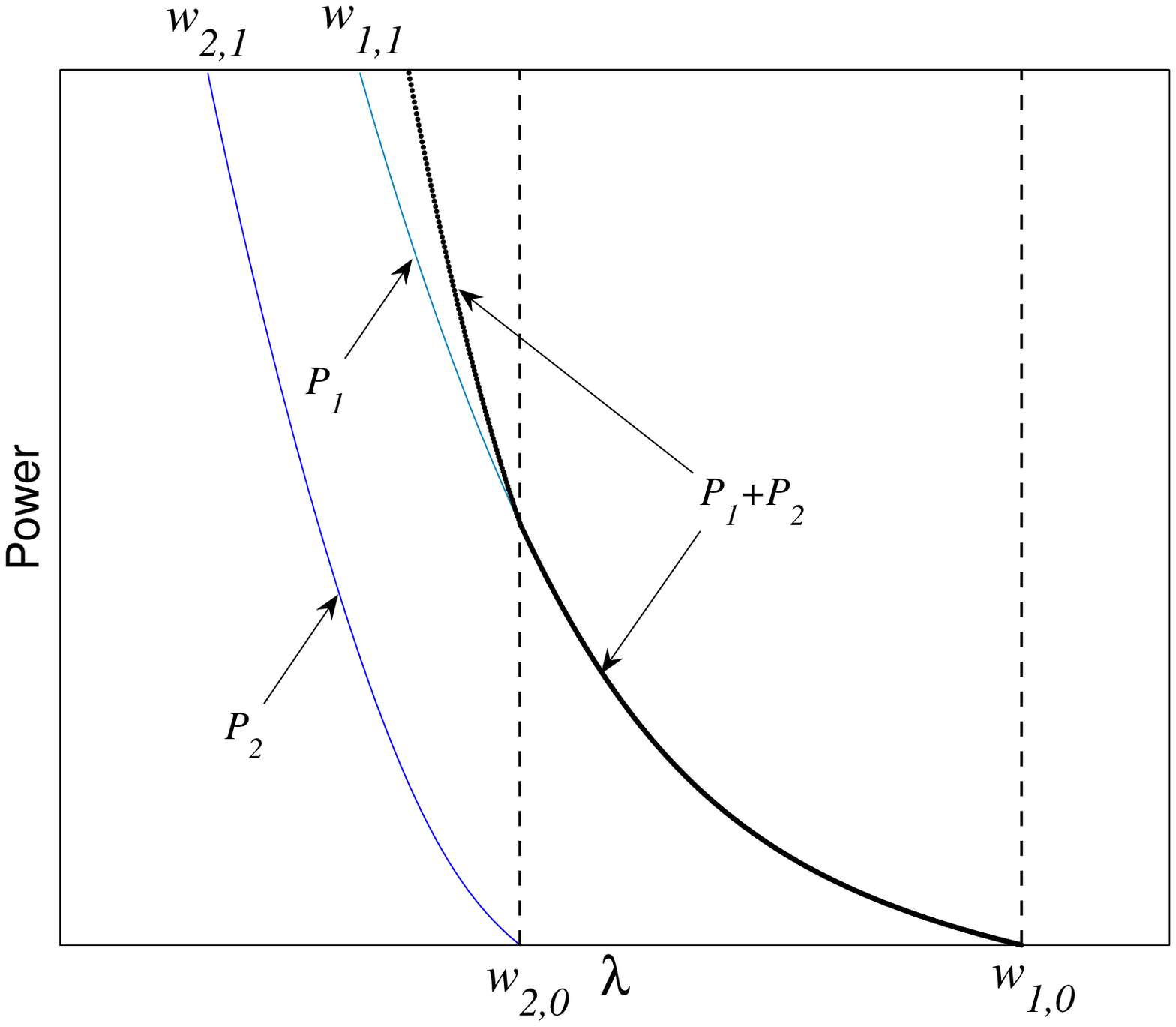}
\caption{\label{fig:case1_Pwr_lambda} Power allocation as a
function of $\lambda^*$ when all the sensors are operating in
region $\mathbf{\cal S}$.}
\end{minipage}
\hfill
\begin{minipage}{0.31\linewidth}
\includegraphics[width=\textwidth]{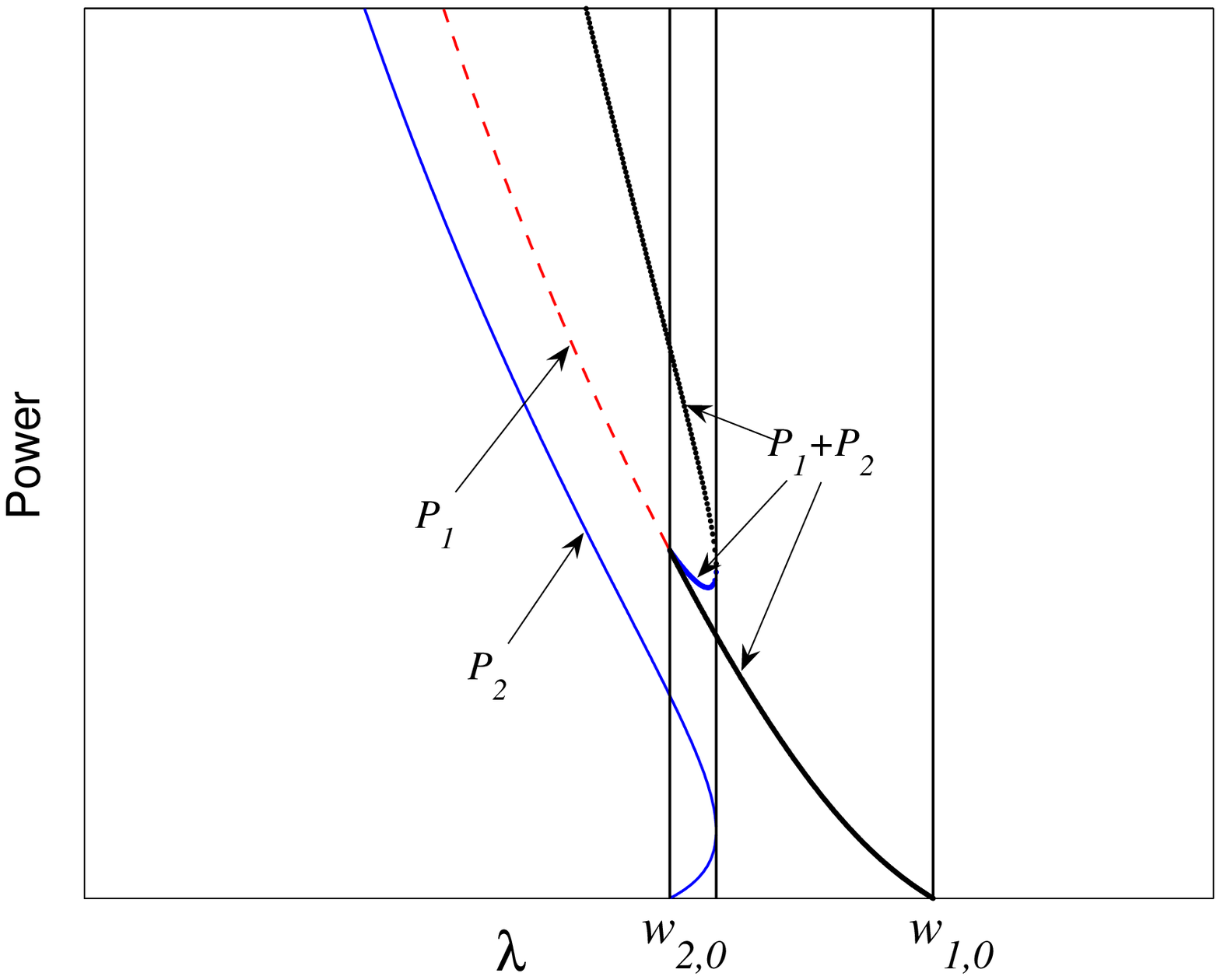}
\caption{\label{fig:case2_Pwr_lambda} Power allocation as a
function of $\lambda^*$ when one or more sensors are not operating
in region $\mathbf{\cal S}$.}
\end{minipage}
\hfill
\end{figure*}

To derive the algorithm, we will use the technique of Lagrange
multipliers \cite{Bertsekas99book}\cite{Boyd_book}. The Lagrangian
associated with the constrained optimization problem in
\eqref{eqn:J approx optimization special case} is
\begin{align}
\label{eqn:L def} &L(P_1,\cdots,P_K,\lambda) \notag \\
& =
\sum_{j=1}^K \left[\frac{\sigma_j^2+\alpha_F(j)P_j}
{\sigma_j^2+\beta_F(j)P_j}+\frac{\sigma_j^2+\alpha_D(j)P_j}{\sigma_j^2+\beta_D(j)P_j}
\right] \nonumber \\
& \qquad -\lambda\left(\sum_{j=1}^KP_j-P_{\rm tot}\right) +
\sum_{j=1}^K\nu_jP_j \notag \\
& \qquad \qquad - \sum_{j=1}^K\eta_j(P_j-P_{\rm max}(j)),
\end{align}
where $\lambda$, $\{\nu_j\}_{j=1}^K$, and $\{\eta_j\}_{j=1}^K$ are
Lagrange multipliers. The Karush-Kuhn-Tucker necessary conditions
for optimality \cite{Boyd_book} are
\begin{align}
\label{eqn:KKT special case} \frac{\partial}{\partial
P_j}J(P_1,\cdots,P_K)\Big |_{P^*_j} -\lambda^*+\nu^*_j
-\eta^*_j&=0, \end{align}
\begin{align}
&\lambda^* > 0, \quad {\rm if}\;\;\sum_{j=1}^KP^*_j=P_{\rm tot},\\
&\lambda^* = 0, \quad {\rm if}\;\;\sum_{j=1}^KP^*_j<P_{\rm
tot},\label{eqn:KKT lambda}\\
&\nu^*_j>0, \quad {\rm if}\;\;P^*_j=0,\\
&\nu^*_j=0, \quad {\rm if}\;\;P^*_j>0,\\[-3mm]
\intertext{and}\notag \\[-8mm]
&\eta^*_j>0, \quad {\rm if}\;\;P^*_j=P_{\rm max}(j),\\
&\eta^*_j=0, \quad {\rm if}\;\;P^*_j<P_{\rm max}(j).
\end{align}
All variables with superscript ``$*$'' are at their optimal
values. Since the optimal solution is always on the total power
constraint boundary as indicated by Lemma \ref{lem: special case J
non decreasing}, \eqref{eqn:KKT lambda} is inapplicable except in
the trivial case when $\sum_{j=1}^KP_{\rm max}(j)<P_{\rm tot}$ and
all the sensors just transmit at full power. So, in other words,
we consider that the total power constraint is always active
(meaning $\sum_{j=1}^KP^*_j = P_{\rm tot}$) and $\lambda^*$ is
always positive. Similarly, $\nu^*_j$ (or $\eta^*_j$) is positive
only when the constraint $P^*_j\ge0$ (or $P^*_j\le P_{\rm
max}(j)$) is active.

\eqref{eqn:KKT special case} is the key equation to solve. For
each fixed value of $\lambda^*$, we can solve \eqref{eqn:KKT
special case} to obtain the corresponding $P_j^*$, $\nu_j^*$, and
$\eta_j^*$. We can then calculate the corresponding
$\sum_{j=1}^KP^*_j$. The goal here is to find a $\lambda^*$ such
that $\sum_{j=1}^KP^*_j=P_{\rm tot}$.

Substituting \eqref{eqn:1st derivative of J(P)} into
\eqref{eqn:KKT special case}, we have
\begin{align} \label{eqn:KKT sub1}
&\frac{(\alpha_F(j)-\beta_F(j))\sigma^2g_j}
{(\sigma^2+\beta_F(j)g_jP_j)^2}
+\frac{(\alpha_D(j)-\beta_D(j))\sigma^2g_j}
{(\sigma^2+\beta_D(j)g_jP_j)^2} \notag \\
& \qquad \qquad \qquad \qquad -\lambda^*+\nu^*_j -\eta^*_j=0.
\end{align}
Let us define
\begin{align} \label{eqn:def w_j0}
w_{j,0}&\eqdef \frac{\partial}{\partial P_j}J(P_1,\cdots,P_K)\Big
|_{P^*_j=0} \notag \\
&=g_j(\alpha_F(j)-\beta_F(j))+\alpha_D(j)-\beta_D(j))/\sigma^2, \\[-2mm]
\intertext{and}\notag \\[-8mm] w_{j,1}&\eqdef \frac{\partial}{\partial
P_j}J(P_1,\cdots,P_K)\Big
|_{P^*_j=P_{\rm max}(j)} \nonumber \\
&=\frac{(\alpha_F(j)-\beta_F(j))\sigma^2g_j} {(\sigma^2+\beta_F(j)
g_j P_{\rm max}(j))^2} \notag \\
&\qquad \qquad +\frac{(\alpha_D(j)-\beta_D(j))\sigma^2g_j}
{(\sigma^2+\beta_D(j)g_jP_{\rm max}(j))^2}.\label{eqn:def w_j1}
\end{align}
For sensor $j$ operating at $(P_D(j),P_F(j))\in \mathbf{\cal S}$,
we can see that, when $\lambda^*>w_{j,0}$, we have $P_j^*=0$,
$\eta^*_j=0$, and $\nu^*_j=\lambda^*-w_{j,0}$. Sensor $j$ starts
to get positive power allocation $P_j^*>0$ when $\lambda^* <
w_{j,0}$, and in this case the constraint $0\le P^*_j\le P_{\rm
max}(j)$ is inactive ($\nu^*_j=\eta_j^*=0$). $\lambda^*$ is
monotonically decreasing with increasing $P_j^*$, as long as
$w_{j,1}<\lambda^* < w_{j,0}$. This can be easily verified using
Lemma \ref{lem: special case J second derivative} since
$\frac{\partial^2}{\partial P_j^2}J(P_1,\cdots,P_K)$ is always
non-positive for $(P_D(j),P_F(j))\in \mathbf{\cal S}$. When
$\lambda^* < w_{j,1}$, $P_j^*$ is now ``clamped'' at $P_{\rm
max}(j)$, so we have $\nu^*_j=0$ and $\eta^*_j=w_{j,1}-\lambda^* >
0$.

This case is depicted as the dashed line in Figure
\ref{fig:lambda_Pj}. Regardless of the value of $\lambda^*$, we
can easily verify that there is always a one-to-one mapping
(through \eqref{eqn:KKT special case} or \eqref{eqn:KKT sub1})
between $P_j^*$ and $\lambda^*$, and $P_j^*$ is nondecreasing with
decreasing $\lambda^*$.

We have the following observations based on the above analysis.
(1) If all the sensors operate at $(P_D(j),P_F(j))\in \mathbf{\cal
S}$, $j=1,\cdots,K$, there is a one-to-one mapping between
$\sum_{j=1}^KP^*_j$ and $\lambda^*$, and $\sum_{j=1}^KP^*_j$ is
nondecreasing with decreasing $\lambda$. (2) The sensors get
positive power allocation with increasing power budget (hence
decreasing $\lambda^*$) in a sequential fashion, and it is
determined by $w_{j,0}$.

The above observations with a two-sensor case are illustrated in
Figure \ref{fig:case1_Pwr_lambda}. Based on the observations, the
solution can be found through a ``weighted waterfilling''
procedure, specified by the following algorithm. Initially the
sensors send their local detection quality
$\{P_D(j),P_F(j))\}_{j=1}^K$ to the FC, and then the algorithm is
executed at the FC.

\noindent \textbf{Algorithm 1:} {\em
\begin{itemize}

\item [(1)] The FC estimates the channel power gain
$\{g_j\}_{j=1}^K$ of the sensors and the noise variance
$\sigma^2$.

\item [(2)] The FC calculates $\{w_{j,0}\}_{j=1}^K$ using
\eqref{eqn:def w_j0} and ranks them such that $w_{j,0} \ge
w_{j+1,0}$. The FC also solves for $\{w_{j,1}\}_{j=1}^K$ using
\eqref{eqn:def w_j1}. Then the FC calculates the power allocations
$\{P^*_{k,w_{j,0}}\}_{k=1}^K$ with $\lambda^*=w_{j,0}$ for each
$j$ using \eqref{eqn:KKT sub1}.

\item[(3)] The FC finds the largest $j'$ such that
$\sum_{k=1}^KP^*_{k,w_{j',0}}\le P_{\rm tot}$, and assigns
$w_a=w_{j',0}$ and $w_b=w_{j'+1,0}$. If $j'=K$, the FC sets
$w_b=0$. The rest of the algorithm conducts a simple line search
on $\lambda^*$ between $w_a$ and $w_b$ such that
$\sum_{k=1}^KP^*_{k,\lambda^*}= P_{\rm tot}$.

\item[(4)] If $\sum_{k=1}^KP^*_{k,w_b}-\sum_{k=1}^KP^*_{k,w_a}\ge
\epsilon$, where $\epsilon$ is a small positive number, the
algorithm goes to (5). Otherwise, the FC stops the algorithm and
broadcasts the desired power allocations $\{P^*_{k,w_a}\}_{k=1}^K$
to the sensors.

\item[(5)] The FC sets $w_c=(w_a+w_b)/2$, following bisection
rule. If $\sum_{k=1}^KP^*_{k,w_c}\le P_{\rm tot}$, the FC sets
$w_a=w_c$. Otherwise, the FC sets $w_b=w_c$. The algorithm goes to
(4).
\end{itemize}}

Algorithm 1 can be easily seen to converge because of the
monotonicity between total power budget $P_{\rm tot}$ and
$\lambda^*$. And the simple line search of $\lambda^*$ between
$w_a$ and $w_b$ should converge very quickly
\cite{Bertsekas99book}.

If sensor $j$ operates at $(P_D(j),P_F(j))\not\in \mathbf{\cal
S}$, $\lambda^*$ is monotonically increasing with $P_j^*$ when
$P_j^*$ is small and is monotonically decreasing with $P_j^*$ when
$P_j^*$ grows larger. This is because $C_0$ is negative and $C_1$,
$C_2$, and $C_3$ are nonnegative in \eqref{eqn:2nd derivative of
J(P)}. Thus $\frac{\partial^2}{\partial P_j^2}J(P_1,\cdots,P_K)
\big|_{P^*_j=0}$ is positive, and will eventually become negative
with $P^*_j$ sufficiently large. Therefore,
$\frac{\partial}{\partial P_j}J(P_1,\cdots,P_K)$ has a single
local maximum at some $P^*_j>0$, and it is possible that
\eqref{eqn:KKT sub1} has two solutions for $P^*_j$ for a single
$\lambda^*$. This case is also shown in Figure
\ref{fig:lambda_Pj}.

If one or more sensors operate at $(P_D(j),P_F(j))\not\in
\mathbf{\cal S}$, the monotonicity and one-to-one mapping between
$\lambda^*$ and $\sum_{j=1}^KP^*_j$ may be invalid, as shown in
Figure \ref{fig:case2_Pwr_lambda}. Therefore, the computationally
efficient Algorithm 1 does not work for this case. The solution
can still be obtained from general constrained optimization
techniques, such as the interior point method
\cite{Bertsekas99book}\cite{Boyd_book}.

\section{Simulations}
\label{sec:simulations} In this section, numerical results are
provided to illustrate the power allocation scheme developed in
this paper. In the simulations, we consider the following
settings. There are $K$ sensors scattered around an FC and the
distances from the sensors to the FC are $\{d_k\}_{k=1}^K$. The
pathloss of signal power at the FC from sensor $k$ follows the
Motley-Keenan pathloss model (expressed in dB) without the wall
and floor attenuation factor \cite{Motley Keenan PL 1988}:
\begin{align}
PL_k=PL_0+10n{\rm log}_{10}(d_k/d_0) \label{eqn:PL model}
\end{align}
where $PL_0$ is a constant set to 55 dB, and $d_0$ is also a
constant set to 1 m in the simulations. Here, $n$ is the pathloss
exponent, which is set to $2$ for free space propagation. The
channel power gain for sensor $k$ is $g_k=-PL_k$ in dB. The noise
variance at the FC is $\sigma^2=-70$ dBm, and we assume the noise
covariance matrix is ${\bf R}=\sigma^2{\bf I}_k$. The maximum
transmitting power of each sensor is $P_{\rm max}=2$ mW (3 dBm).
The total power budget in the simulations will be below or equal
to $2K$ mW, otherwise each sensor will just use maximum
transmitting power (a trivial case). All the sensors perform
Neyman-Pearson detection with false alarm probabilities set to
$P_F(k)=0.04$, $k=1,\cdots,K$. The detection probabilities may
vary according to their local observation qualities. The FC also
uses a Neyman-Pearson detector targeting the same false alarm
probability as the local sensors\footnote{The operation points in
terms of targeted false alarm probabilities of the detectors at
the sensors and the FC can be designed to be different from one
another, and the analysis in this paper does not require the false
alarm probabilities to be the same. The optimal design of the
operating points of the detectors is beyond the scope of this
paper, and we use the same false alarm probability for the sake of
simplicity.}, i.e., $P_{F,FC}=0.04$.

We will investigate three scenarios: 1) two sensors with
orthogonal MIMO channels. 2) two sensors with non-orthogonal MIMO
channels. 3) ten sensors with orthogonal MIMO channels.

\subsection{Two Sensors with orthogonal channels}
\label{subsec:two sensors with orthogonal channels} Two sensors
are located $d_1=2$ m and $d_2=5$ m away from the FC, and
communicate with the FC through orthogonal channels. Channel gains
$g_1$ and $g_2$ are calculated by \eqref{eqn:PL model}, and are
-61 dB and -69 dB respectively.

We will consider four cases with various local detection quality
combinations.
\begin{enumerate} \item[] Case \ref{subsec:two sensors with
orthogonal channels}1: $P_D(1)=0.1$, $P_D(2)=0.9$. \item[] Case
\ref{subsec:two sensors with orthogonal channels}2: $P_D(1)=0.7$,
$P_D(2)=0.9$. \item[] Case \ref{subsec:two sensors with orthogonal
channels}3: $P_D(1)=0.9$, $P_D(2)=0.9$. \item[] Case
\ref{subsec:two sensors with orthogonal channels}4: $P_D(1)=0.9$,
$P_D(2)=0.7$.
\end{enumerate}
In Case \ref{subsec:two sensors with orthogonal channels}1 one
sensor does not operate in region $\mathbf{\cal S}$, so the
interior point optimization algorithm is used in this case. In
Case \ref{subsec:two sensors with orthogonal channels}2, Case
\ref{subsec:two sensors with orthogonal channels}3, and Case
\ref{subsec:two sensors with orthogonal channels}4, both sensors
operate in region $\mathbf{\cal S}$, thus Algorithm 1 is used. The
total power budget $P_{\rm tot}$ varies from -14 dBm to 6 dBm
(when each sensor transmits at full power 2 mW).

In addition to the proposed power allocation, we also include an
equal power allocation and an equal received SNR power allocation
for comparison. The equal power allocation simply distributes
power equally among the sensors, without considering channel or
local detection quality. The equal received SNR power allocation
considers channel quality only and distributes power among sensors
in such a way that the received signals from the sensors have the
same SNR.

\begin{figure}[tb]
\includegraphics[width=0.38\textwidth]{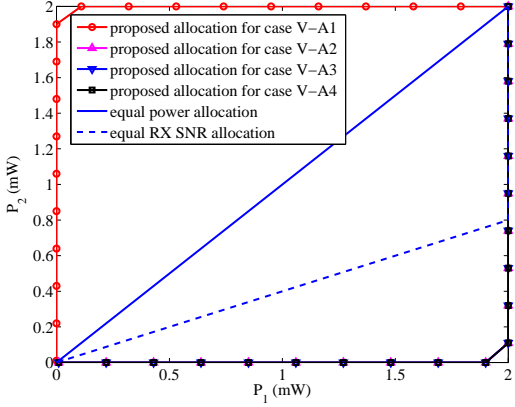}
\caption{\label{fig:two sensor orth channel proposed allocation}
Equal power allocation, equal received SNR allocation, and the
proposed power allocation for the four cases in Section
\ref{subsec:two sensors with orthogonal channels}. (Note that the
curves for $\vartriangle$, $\triangledown$ and $\square$ overlay
in this graph.)}
\end{figure}

Figure \ref{fig:two sensor orth channel proposed allocation} shows
the proposed power allocation as well as the equal power
allocation and the equal received SNR power allocation, we can see
that for Case \ref{subsec:two sensors with orthogonal channels}1
the proposed power allocation distributes all the power to sensor
2, until the maximum output power is reached for sensor 2, and
then sensor 1 starts to get positive power allocation. This is
because, although sensor 1 is closer to the FC (hence it has a
better channel), its detection quality is much worse than that of
sensor 2. For Case \ref{subsec:two sensors with orthogonal
channels}2, the detection quality of sensor 1 is still worse than
that of sensor 2, but the gap is small enough for sensor 1's
better communication channel to show difference. The proposed
allocation distributes all the power to sensor 1, until the
maximum output power is reached for sensor 1, and then sensor 2
starts to get power allocation. For Case \ref{subsec:two sensors
with orthogonal channels}3 and Case \ref{subsec:two sensors with
orthogonal channels}4, sensor 1 has a better communication channel
and equal or better local detection quality, so it is not
surprising to see that the proposed allocation distributes power
to sensor 1 as much as possible, and these two cases have the same
power allocation as Case \ref{subsec:two sensors with orthogonal
channels}2. The waterfilling effect of the proposed power
allocation is obvious in this scenario. The equal power allocation
and the equal received SNR allocation do not change between the
four cases because they are not affected by the local detection
quality.

\begin{figure}[tb]
\includegraphics[width=0.38\textwidth]{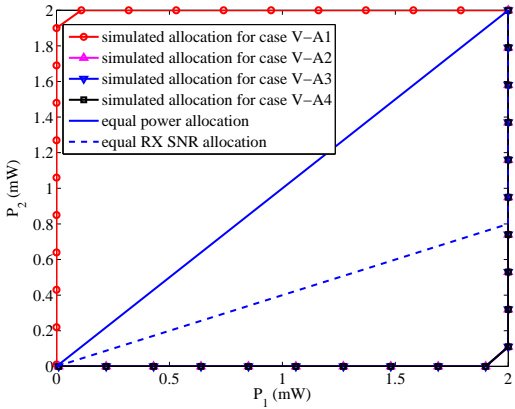}
\caption{\label{fig:two sensor orth channel simu allocation} Equal
power allocation, equal received SNR allocation, and the simulated
optimal power allocation for the four cases in Section
\ref{subsec:two sensors with orthogonal channels}. (Note that the
curves for $\vartriangle$, $\triangledown$ and $\square$ overlay
in this graph.)}
\end{figure}

Recall that the proposed power allocation scheme is based on the
J-divergence instead of detection probability and false alarm
probability. Furthermore, the J-divergence between two Gaussian
mixture distributions, i.e., that of the received signals at the
FC under the two hypotheses, is approximated in this optimization
by the J-divergence between two Gaussian distributions, with the
same means and covariance matrices as the Gaussian mixtures. We
next show the quality of this approximation.

Figure \ref{fig:two sensor orth channel simu allocation} shows the
optimal power allocation found by simulations. The FC uses a
Neyman-Pearson detector based on the likelihood ratio of the
received signal ${\bf y}$. The optimal power allocation is the one
that produces the highest $P_{D,FC}$ for a given total power
budget. The optimal power allocation is found by a brute-force
grid search in a two dimensional space of all possible power
allocations. For each possible power allocation point, $2\times
10^4$ Monte Carlo runs are used to provide the corresponding
$P_{D,FC}$.

We can see that the simulated optimal power allocations in Figure
\ref{fig:two sensor orth channel simu allocation} perfectly match
the proposed power allocations in Figure \ref{fig:two sensor orth
channel proposed allocation}.

The contours of the approximated J-divergence (used as the
objective function to develop the proposed power allocation) and
the simulated $P_{D,FC}$ for Case \ref{subsec:two sensors with
orthogonal channels}3 are plotted in Figure \ref{fig:two sensor
orth channel approx J contour case 3} and Figure \ref{fig:two
sensor orth channel simu PD contour case 3}. The contours for the
other cases are similar and are omitted due to limited space. We
can see from Figure \ref{fig:two sensor orth channel approx J
contour case 3} and Figure \ref{fig:two sensor orth channel simu
PD contour case 3} that the two contours match each other well at
any power allocation point in this scenario, including the case in
which either sensor transmits at full power (2 mW or equivalently
3 dBm). When sensor 1 transmits at full power (3 dBm), the
corresponding received SNR is about 12 dB. So we can see the
approximation works well in this scenario even with received SNR
as high as 12 dB. This is the reason for the perfect match between
the proposed allocation and the simulated optimal allocation.

\begin{figure*}[tb]
\begin{minipage}{0.32\linewidth}
\includegraphics[width=\textwidth]{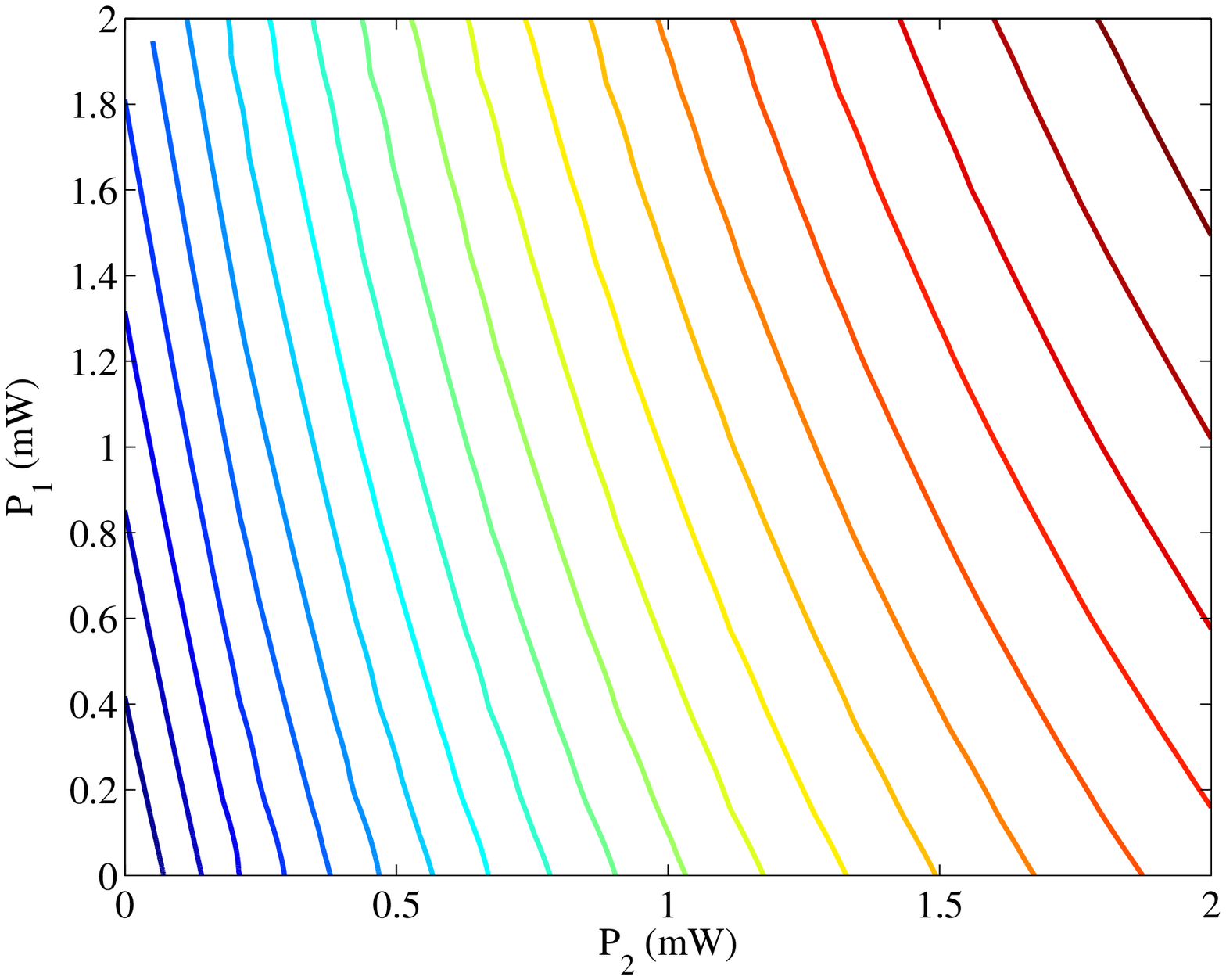}
\caption{\label{fig:two sensor orth channel approx J contour case
3} The contour of the approximated J-divergence objective function
for Case \ref{subsec:two sensors with orthogonal channels}3.}
\end{minipage}
\hfill
\begin{minipage}{0.32\linewidth}
\includegraphics[width=\textwidth]{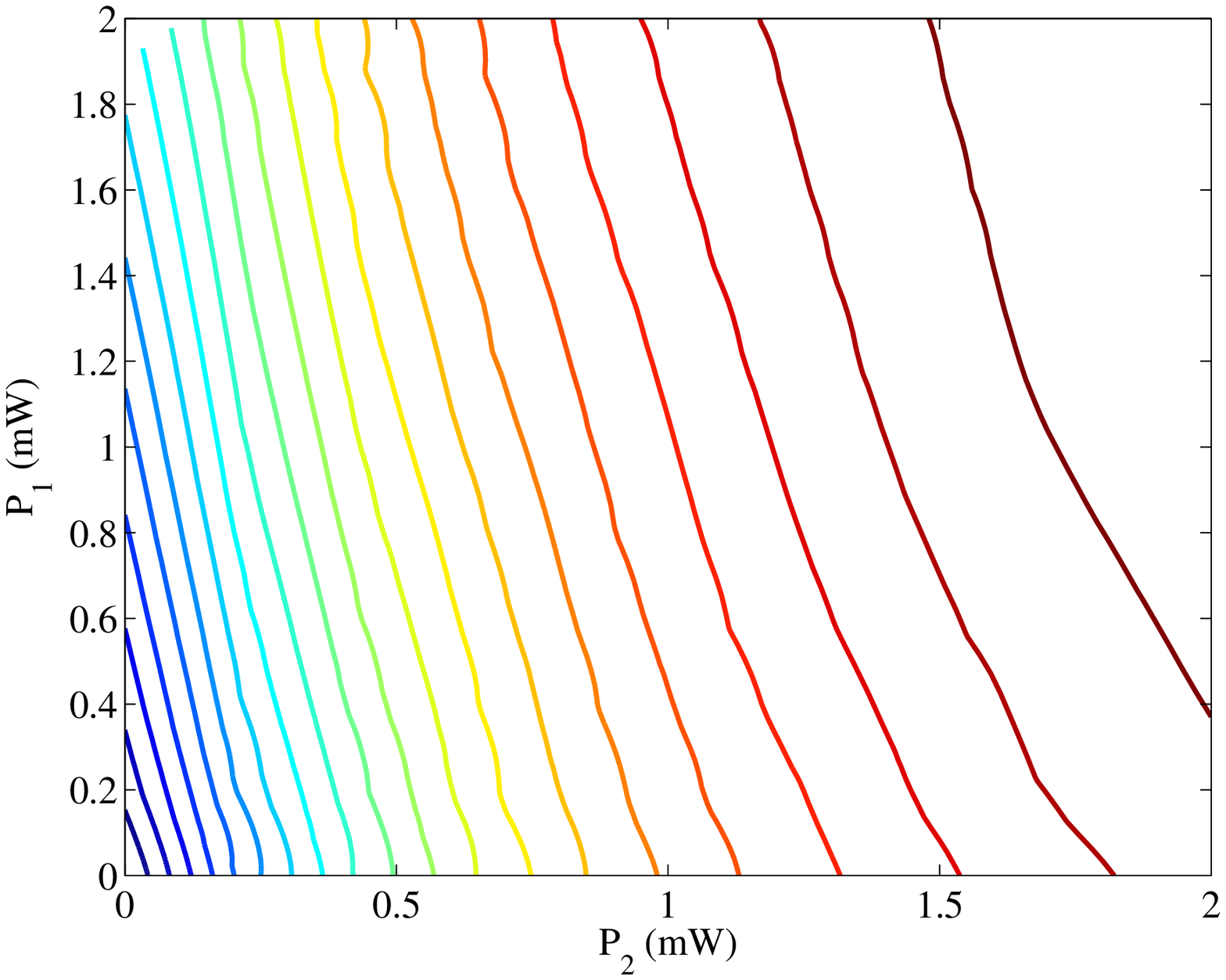}
\caption{\label{fig:two sensor orth channel simu PD contour case
3} The contour of the simulated $P_{D,FC}$ for Case
\ref{subsec:two sensors with orthogonal channels}3.}
\end{minipage}
\hfill
\begin{minipage}{0.32\linewidth}
\includegraphics[width=\textwidth]{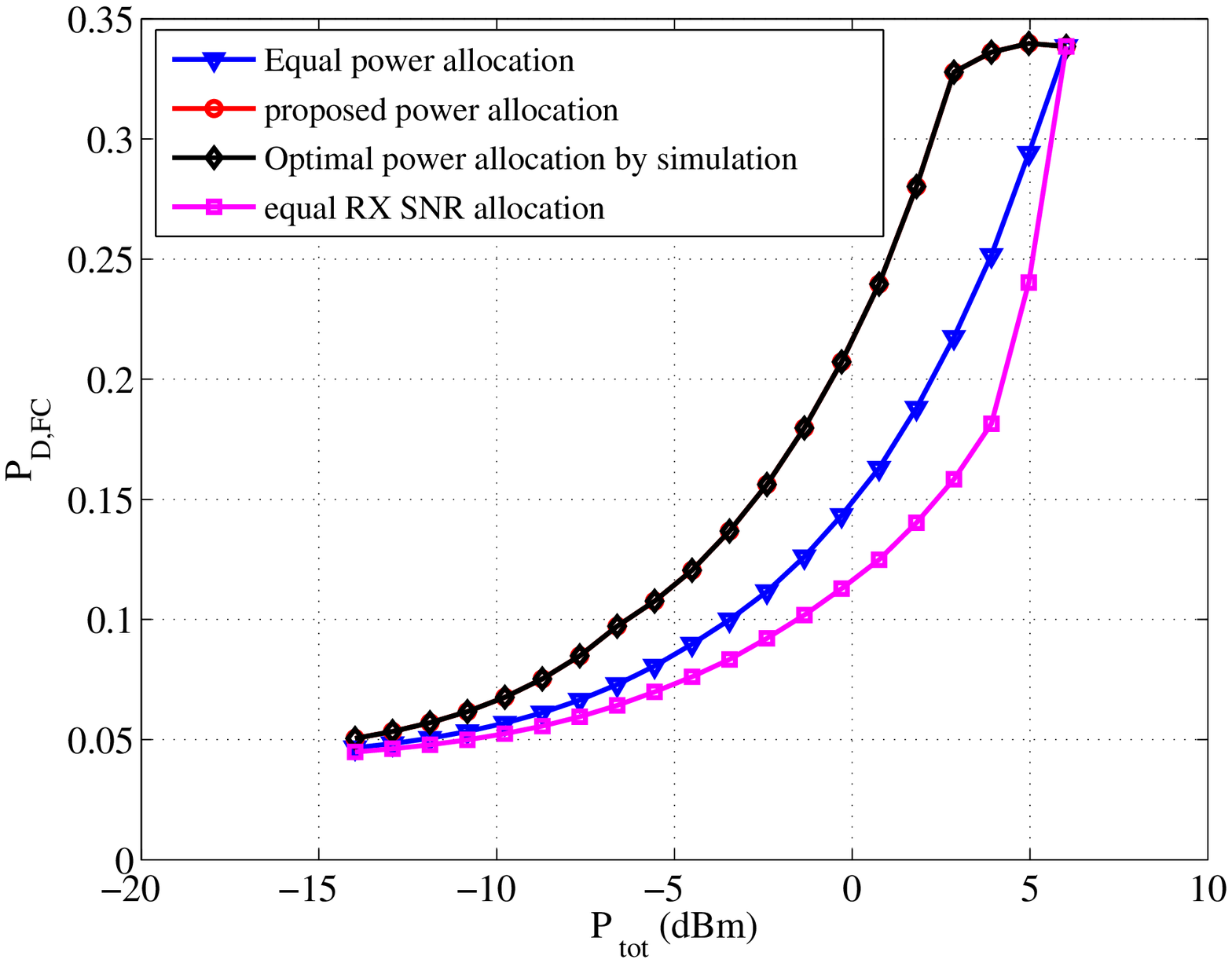}
\caption{\label{fig:two sensor orth channel PDFC vs Ptot case 1}
The FC detection probability $P_{D,FC}$ as a function of $P_{\rm
tot}$ of Case \ref{subsec:two sensors with orthogonal channels}1.}
\end{minipage}
\hfill
\end{figure*}

\begin{figure*}[tb]
\begin{minipage}{0.32\linewidth}
\includegraphics[width=\textwidth]{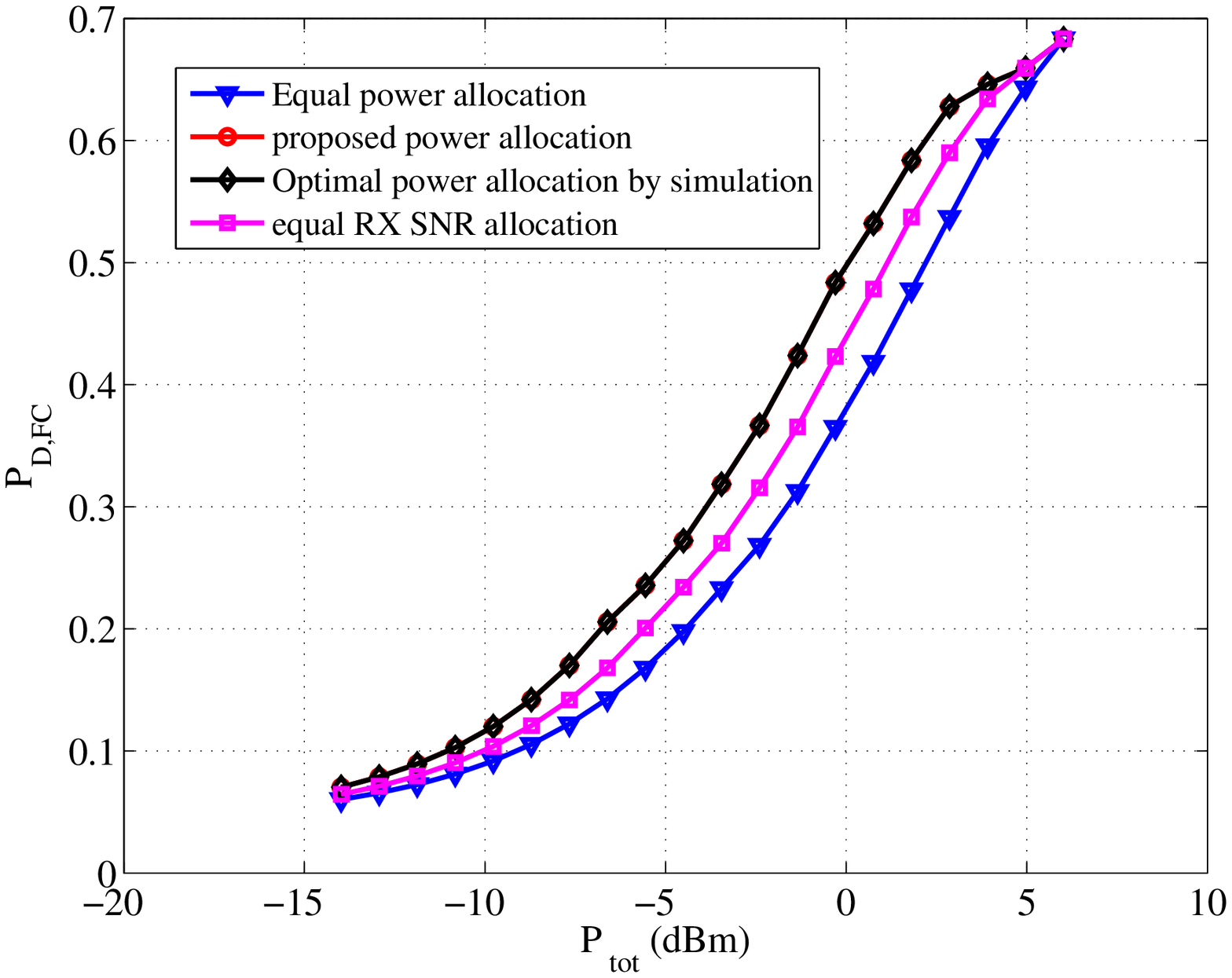}
\caption{\label{fig:two sensor orth channel PDFC vs Ptot case 2}
The FC detection probability $P_{D,FC}$ as a function of $P_{\rm
tot}$ of Case \ref{subsec:two sensors with orthogonal channels}2.}
\end{minipage}
\hfill
\begin{minipage}{0.32\linewidth}
\includegraphics[width=\textwidth]{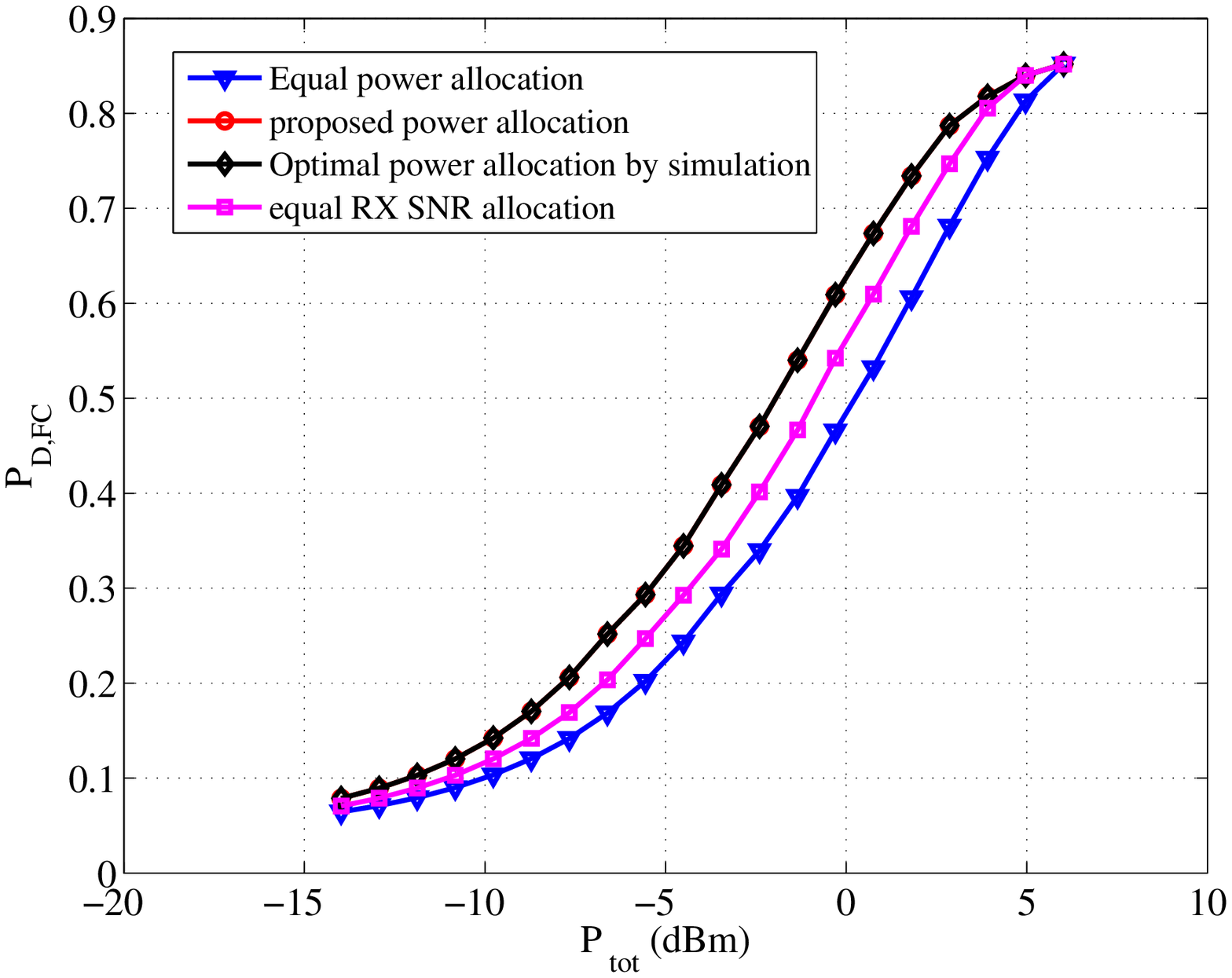}
\caption{\label{fig:two sensor orth channel PDFC vs Ptot case 3}
The FC detection probability $P_{D,FC}$ as a function of $P_{\rm
tot}$ of Case \ref{subsec:two sensors with orthogonal channels}3.}
\end{minipage}
\hfill
\begin{minipage}{0.32\linewidth}
\includegraphics[width=\textwidth]{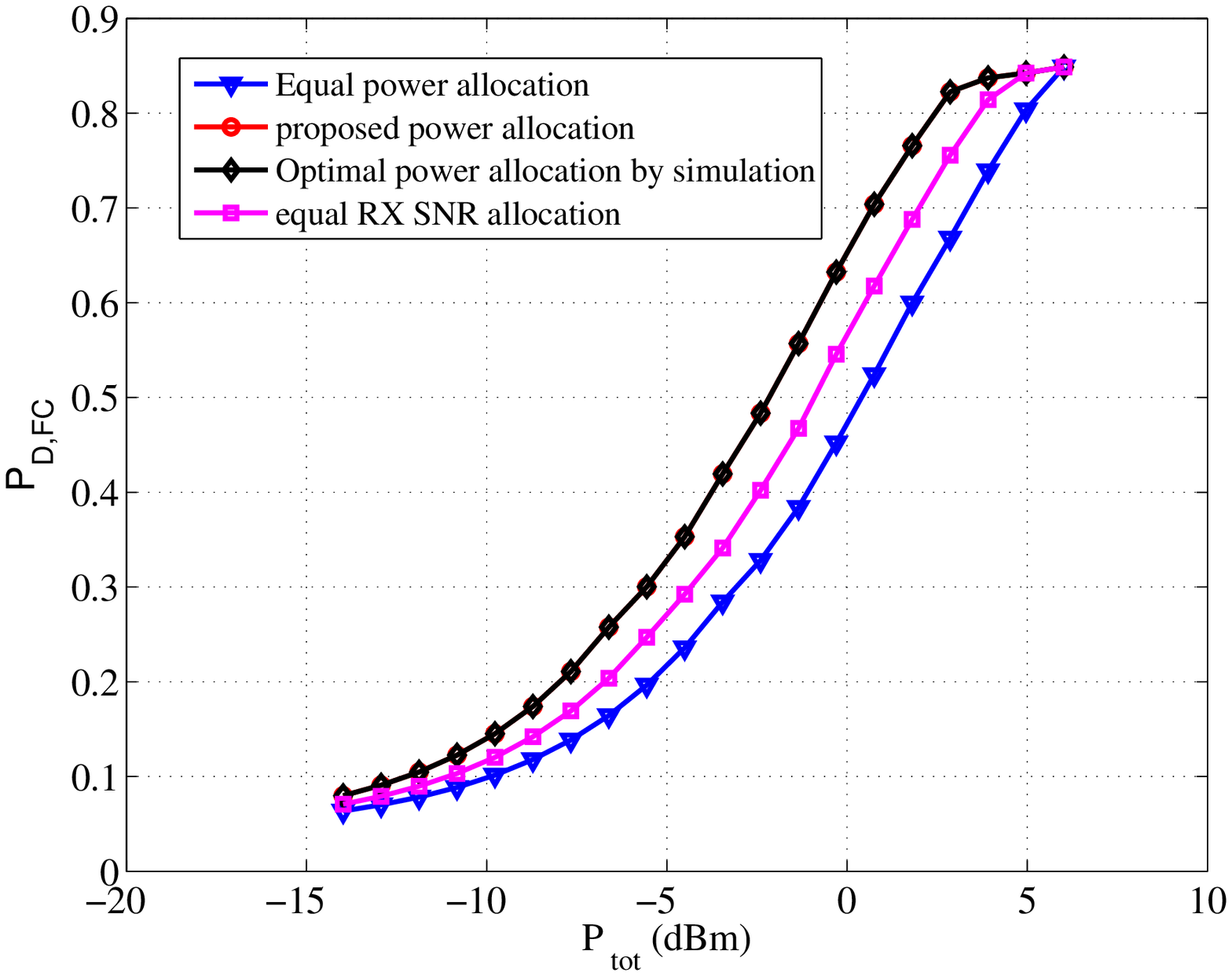}
\caption{\label{fig:two sensor orth channel PDFC vs Ptot case 4}
The FC detection probability $P_{D,FC}$ as a function of $P_{\rm
tot}$ of Case \ref{subsec:two sensors with orthogonal channels}4.}
\end{minipage}
\hfill
\end{figure*}

In Figure \ref{fig:two sensor orth channel PDFC vs Ptot case
1}-Figure \ref{fig:two sensor orth channel PDFC vs Ptot case 4},
we plot the detection probability at the FC $P_{D,FC}$ as a
function of the total power budget $P_{\rm tot}$ for the four
cases, and for the proposed analytical and simulated optimal
allocation. The performance of the proposed power allocation
matches that of the simulated optimal power allocation very well
in all four cases (the two curves overlay in the four figures),
and it can save almost 3 dB in $P_{\rm tot}$ compared to the equal
power allocation to achieve the same $P_{D,FC}$. The equal
received SNR power allocation considers only the channel quality,
so it performs even worse than the equal power allocation in Case
\ref{subsec:two sensors with orthogonal channels}1, where
optimally sensor 1 with a better channel but much worse detection
quality should use less power than sensor 2. On the other hand, in
Case \ref{subsec:two sensors with orthogonal channels}2 through
Case \ref{subsec:two sensors with orthogonal channels}4, the equal
received SNR allocation's performance is between the proposed
allocation and equal allocation.

\subsection{Two sensors with non-orthogonal MIMO channels}
\label{subsec:two sensors with non-orthogonal channels} The
setting here is similar to the setting in Section \ref{subsec:two
sensors with orthogonal channels}, but the data transmission is
over non-orthogonal channels. The channel matrix is given by
\begin{align}
\label{eqn:simu H} {\bf H}&= \begin{bmatrix} 1 & \rho \\
\rho & 1 \end{bmatrix} \begin{bmatrix} g_1 & 0 \\
0 & g_2 \end{bmatrix}.
\end{align}
$g_1$ and $g_2$ are the same as those in Section \ref{subsec:two
sensors with orthogonal channels}. $\rho=0.2$ is the interference
coefficient. We consider four cases, Case \ref{subsec:two sensors
with non-orthogonal channels}1 through Case \ref{subsec:two
sensors with non-orthogonal channels}4, with exactly the same
local detection quality combinations as those of Case
\ref{subsec:two sensors with orthogonal channels}1 through Case
\ref{subsec:two sensors with orthogonal channels}4. The interior
point optimization algorithm is used to solve the proposed power
allocation for all four cases.

Figure \ref{fig:two sensor nonorth channel proposed allocation}
shows the proposed power allocation as well as the equal power
allocation and the equal received SNR allocation. The major
difference between Figure \ref{fig:two sensor nonorth channel
proposed allocation} and Figure \ref{fig:two sensor orth channel
proposed allocation} in Section \ref{subsec:two sensors with
orthogonal channels} is that the waterfilling effect in Case
\ref{subsec:two sensors with non-orthogonal channels}2 through
Case \ref{subsec:two sensors with non-orthogonal channels}4 is not
as obvious as that in Case \ref{subsec:two sensors with orthogonal
channels}2 through Case \ref{subsec:two sensors with orthogonal
channels}4. The non-orthogonal channel makes the contribution from
the sensors at the FC dependent. So the power allocation is less
extreme in the sense that the ``better'' sensors take all the
power.

\begin{figure}[tb]
\includegraphics[width=0.38\textwidth]{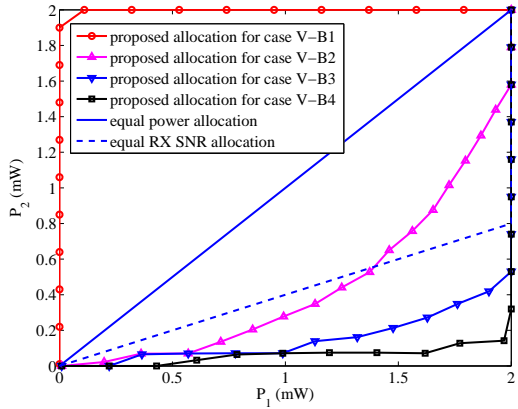}
\caption{\label{fig:two sensor nonorth channel proposed
allocation} Equal power allocation, equal received SNR allocation,
and the proposed power allocation for the four cases in Section
\ref{subsec:two sensors with non-orthogonal channels}. }
\end{figure}
\begin{figure}[tb]
\includegraphics[width=0.38\textwidth]{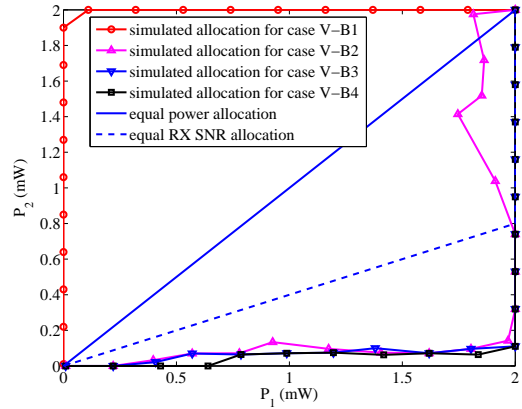}
\caption{\label{fig:two sensor nonorth channel simu allocation}
Equal power allocation, equal received SNR allocation, and the
simulated optimal power allocation for the four cases in Section
\ref{subsec:two sensors with non-orthogonal channels}.}
\end{figure}

Figure \ref{fig:two sensor nonorth channel simu allocation} shows
the optimal power allocation found by simulations for the four
cases. For Case \ref{subsec:two sensors with non-orthogonal
channels}1 and Case \ref{subsec:two sensors with non-orthogonal
channels}4, the simulated optimal power allocations in Figure
\ref{fig:two sensor nonorth channel simu allocation} match the the
proposed power allocations in Figure \ref{fig:two sensor nonorth
channel proposed allocation}. For Case \ref{subsec:two sensors
with non-orthogonal channels}2 and Case \ref{subsec:two sensors
with non-orthogonal channels}3, the simulated optimal power
allocations are different from the proposed power allocation in
higher total power budget region. For these two cases, $P_{D,FC}$
as a function of power allocation is quite ``flat'' in the high
total power region. This can be seen from the wider gaps between
contour lines in Figure \ref{fig:two sensor nonorth channel simu
PD contour case 2}, and from the results in Figure \ref{fig:two
sensor nonorth channel PDFC vs Ptot case 2} and Figure
\ref{fig:two sensor nonorth channel PDFC vs Ptot case 3} that the
performance of the proposed power allocation is still very close
to that of the simulated power allocation. So the artifact of
Monte Carlo trials is more pronounced, and explains the difference
between the proposed power allocations and the simulated optimal
power allocation in the high total power region for Case
\ref{subsec:two sensors with non-orthogonal channels}2 and Case
\ref{subsec:two sensors with non-orthogonal channels}3.

The contours of the approximated J-divergence and the simulated
$P_{D,FC}$ for Case \ref{subsec:two sensors with non-orthogonal
channels}2 are plotted in Figure \ref{fig:two sensor nonorth
channel approx J contour case 2} and Figure \ref{fig:two sensor
nonorth channel simu PD contour case 2}. The two contours match
each other well, but the artifact of Monte Carlo trials in Figure
\ref{fig:two sensor nonorth channel simu PD contour case 2} is
obvious, as discussed above.

\begin{figure*}[tb]
\begin{minipage}{0.32\linewidth}
\includegraphics[width=\textwidth]{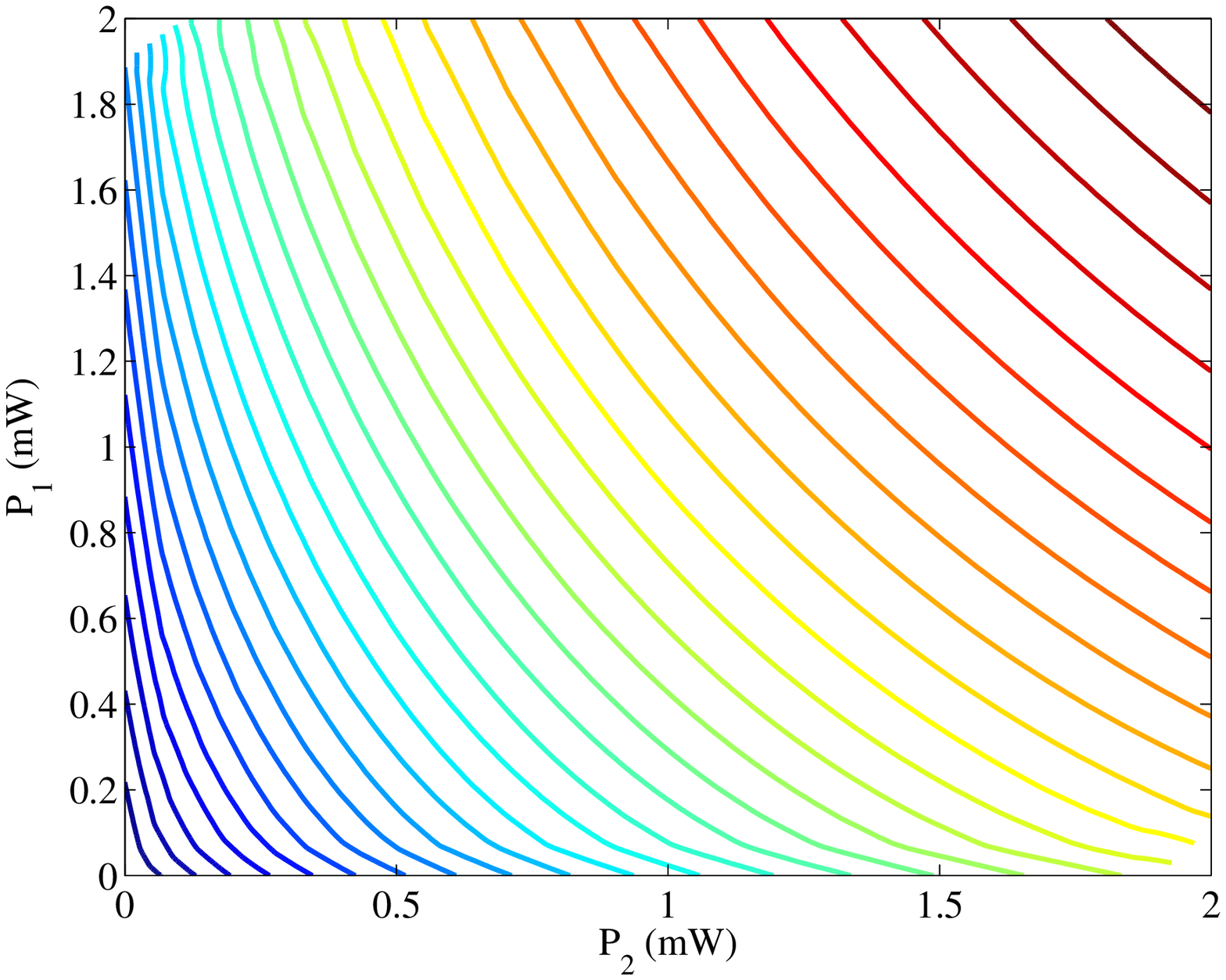}
\caption{\label{fig:two sensor nonorth channel approx J contour
case 2} The contour of the approximated J-divergence objective
function for Case \ref{subsec:two sensors with non-orthogonal
channels}2.}
\end{minipage}
\hfill
\begin{minipage}{0.32\linewidth}
\includegraphics[width=\textwidth]{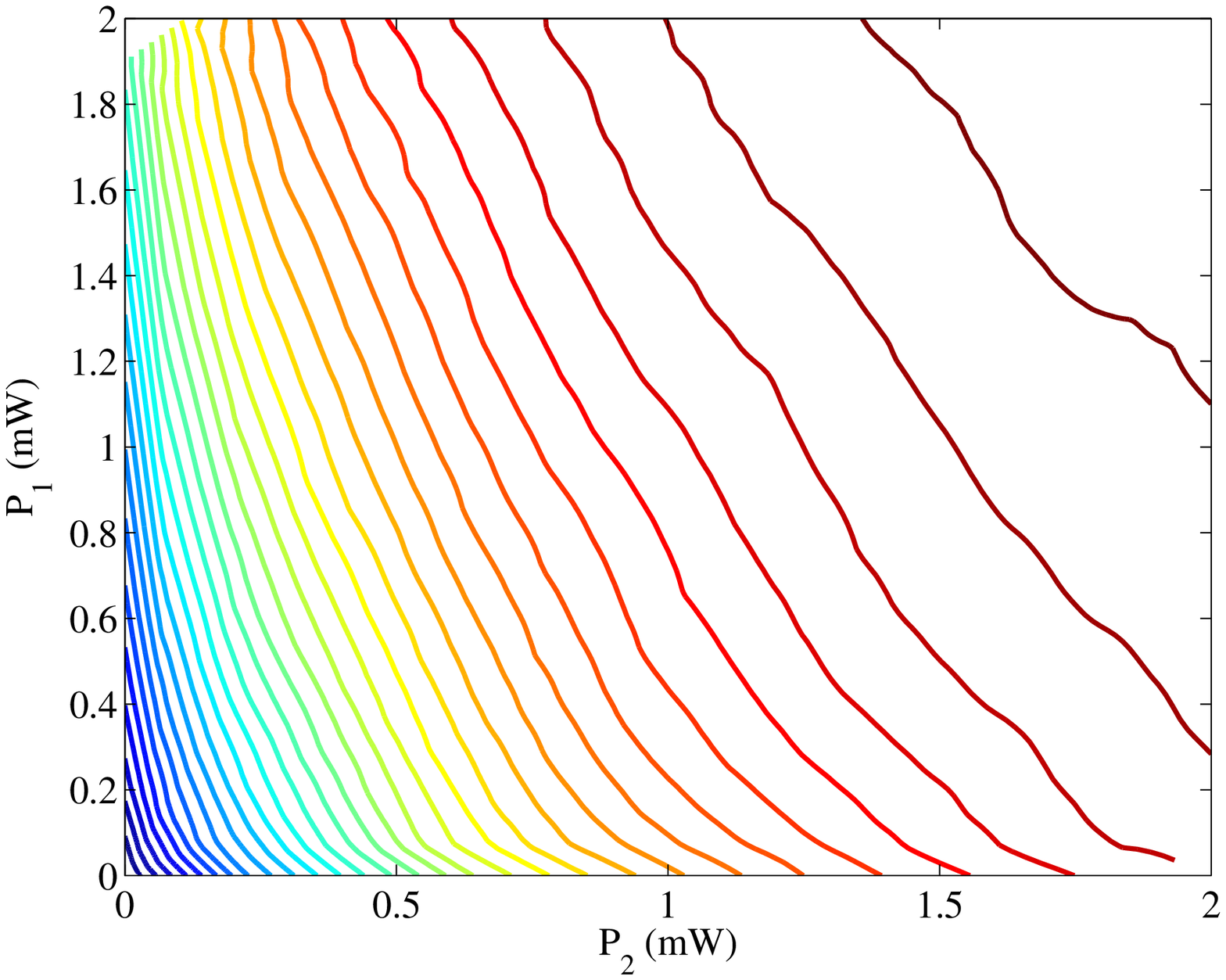}
\caption{\label{fig:two sensor nonorth channel simu PD contour
case 2} The contour of the simulated $P_{D,FC}$ for Case
\ref{subsec:two sensors with non-orthogonal channels}2.}
\end{minipage}
\hfill
\begin{minipage}{0.32\linewidth}
\includegraphics[width=\textwidth]{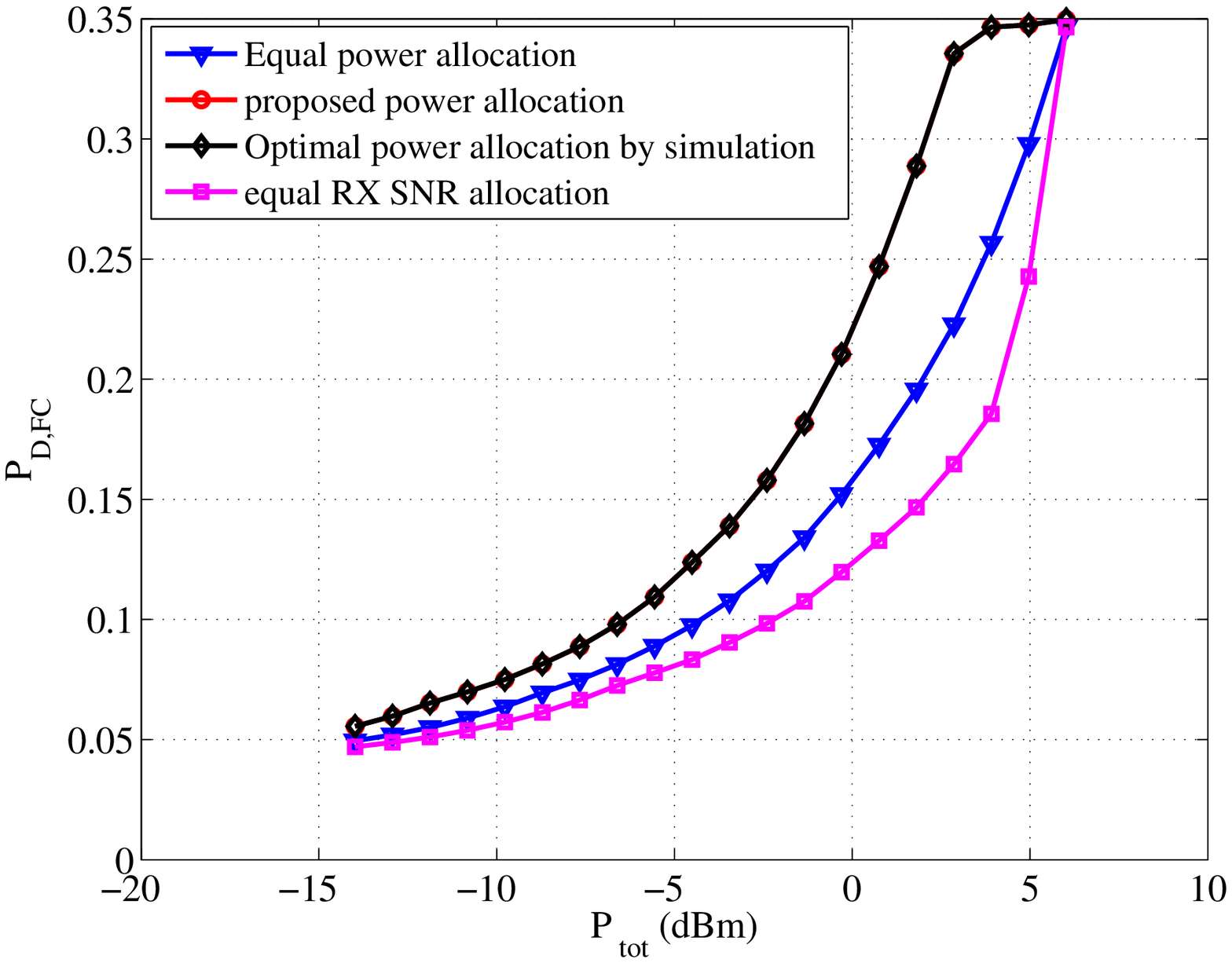}
\caption{\label{fig:two sensor nonorth channel PDFC vs Ptot case
1} The FC detection probability $P_{D,FC}$ as a function of
$P_{\rm tot}$ of Case \ref{subsec:two sensors with non-orthogonal
channels}1.}
\end{minipage}
\hfill
\end{figure*}

In Figure \ref{fig:two sensor nonorth channel PDFC vs Ptot case
1}-Figure \ref{fig:two sensor nonorth channel PDFC vs Ptot case
4}, we plot the detection probability at the FC $P_{D,FC}$ as a
function of the total power budget $P_{\rm tot}$, for the four
cases. The performance of the proposed power allocation matches
that of the simulated optimal power allocation very well in all
four cases (the two curves overlay in the four figures). Compared
to Figure \ref{fig:two sensor orth channel PDFC vs Ptot case
1}-Figure \ref{fig:two sensor orth channel PDFC vs Ptot case 4} in
Section \ref{subsec:two sensors with orthogonal channels}, the
performance gap between the proposed power allocation and the
equal power allocation (as well as the equal received SNR
allocation) is slightly narrower.

\begin{figure*}[tb]
\begin{minipage}{0.32\linewidth}
\includegraphics[width=\textwidth]{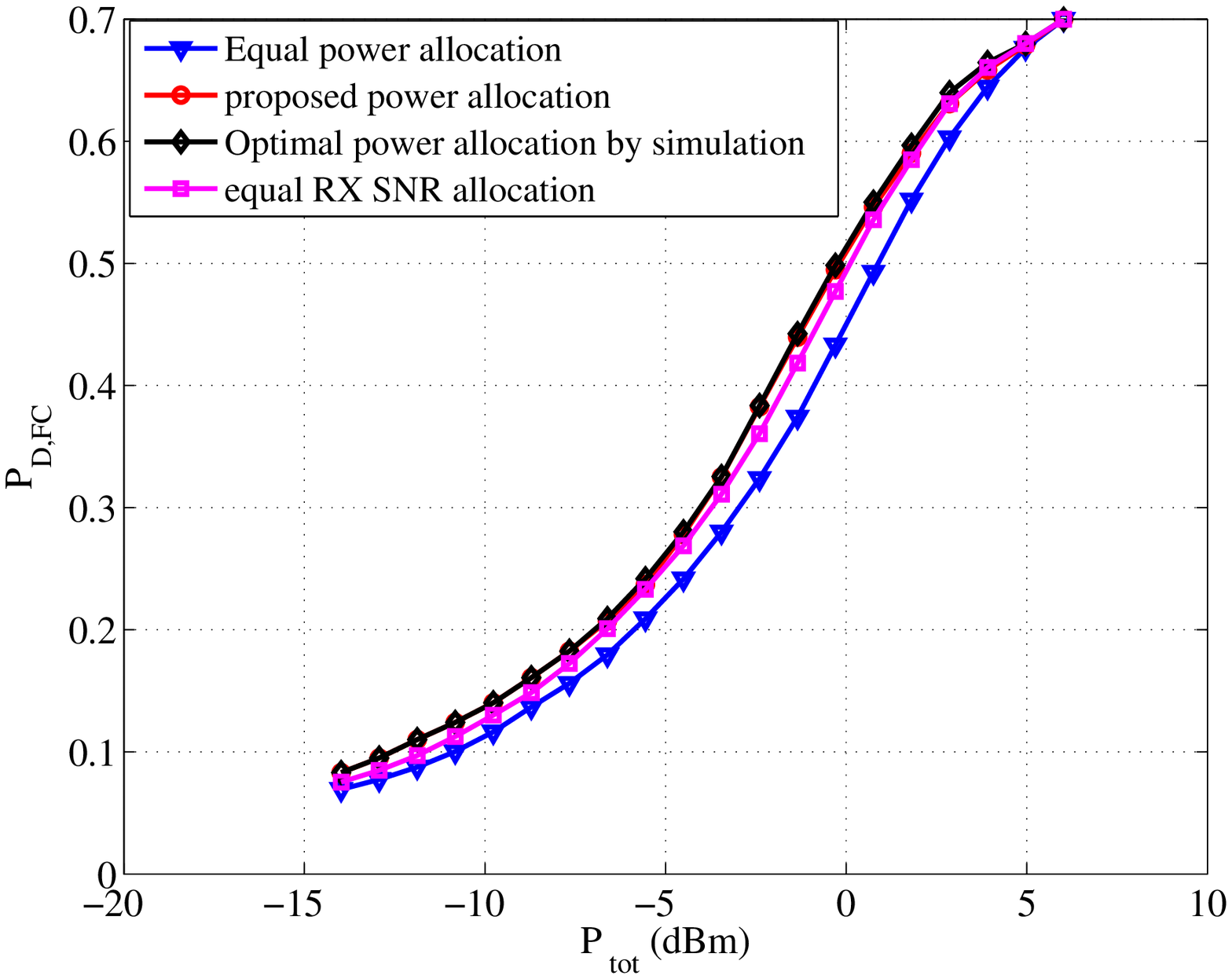}
\caption{\label{fig:two sensor nonorth channel PDFC vs Ptot case
2} The FC detection probability $P_{D,FC}$ as a function of
$P_{\rm tot}$ of Case \ref{subsec:two sensors with non-orthogonal
channels}2.}
\end{minipage}
\hfill
\begin{minipage}{0.32\linewidth}
\includegraphics[width=\textwidth]{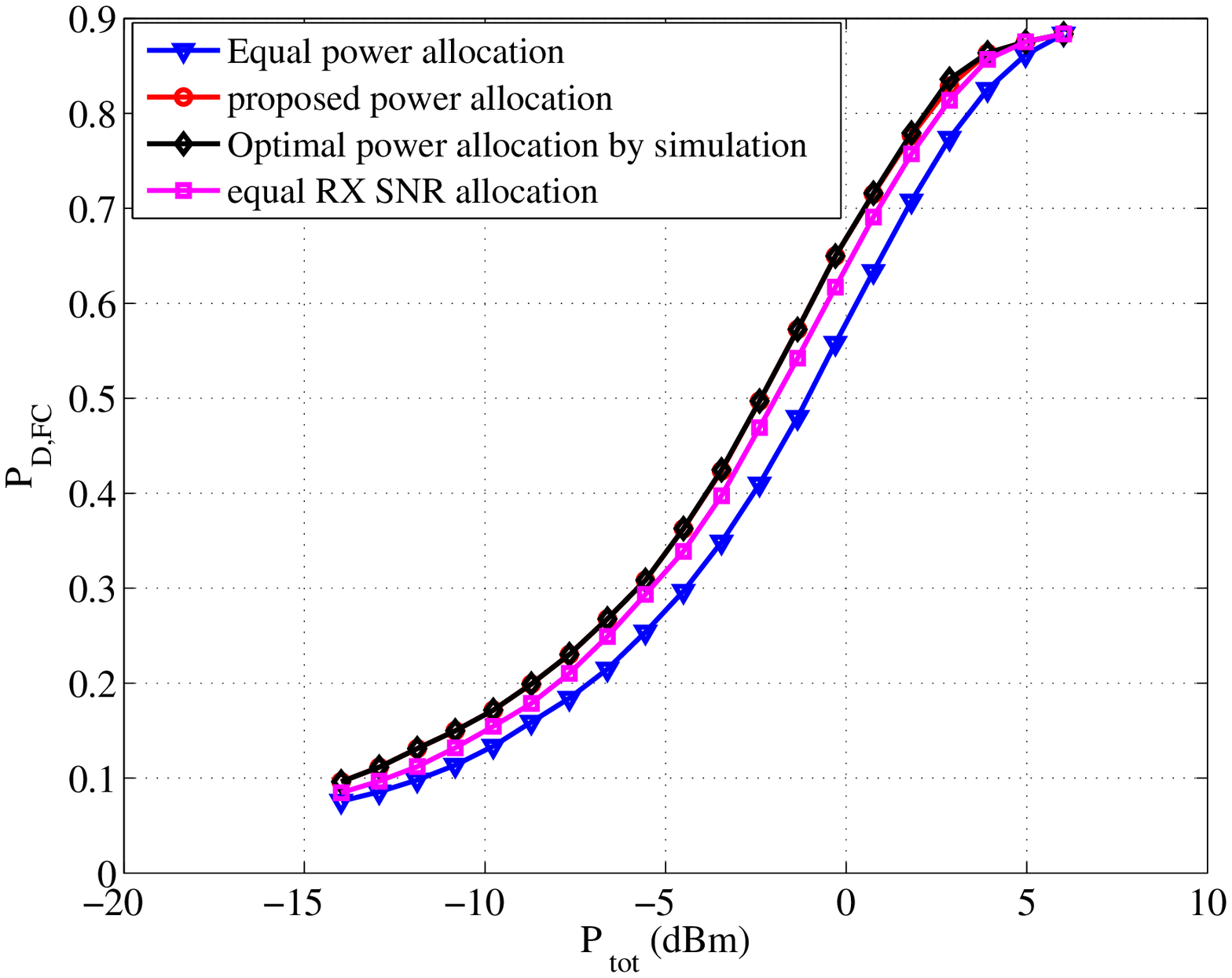}
\caption{\label{fig:two sensor nonorth channel PDFC vs Ptot case
3} The FC detection probability $P_{D,FC}$ as a function of
$P_{\rm tot}$ of Case \ref{subsec:two sensors with non-orthogonal
channels}3.}
\end{minipage}
\hfill
\begin{minipage}{0.32\linewidth}
\includegraphics[width=\textwidth]{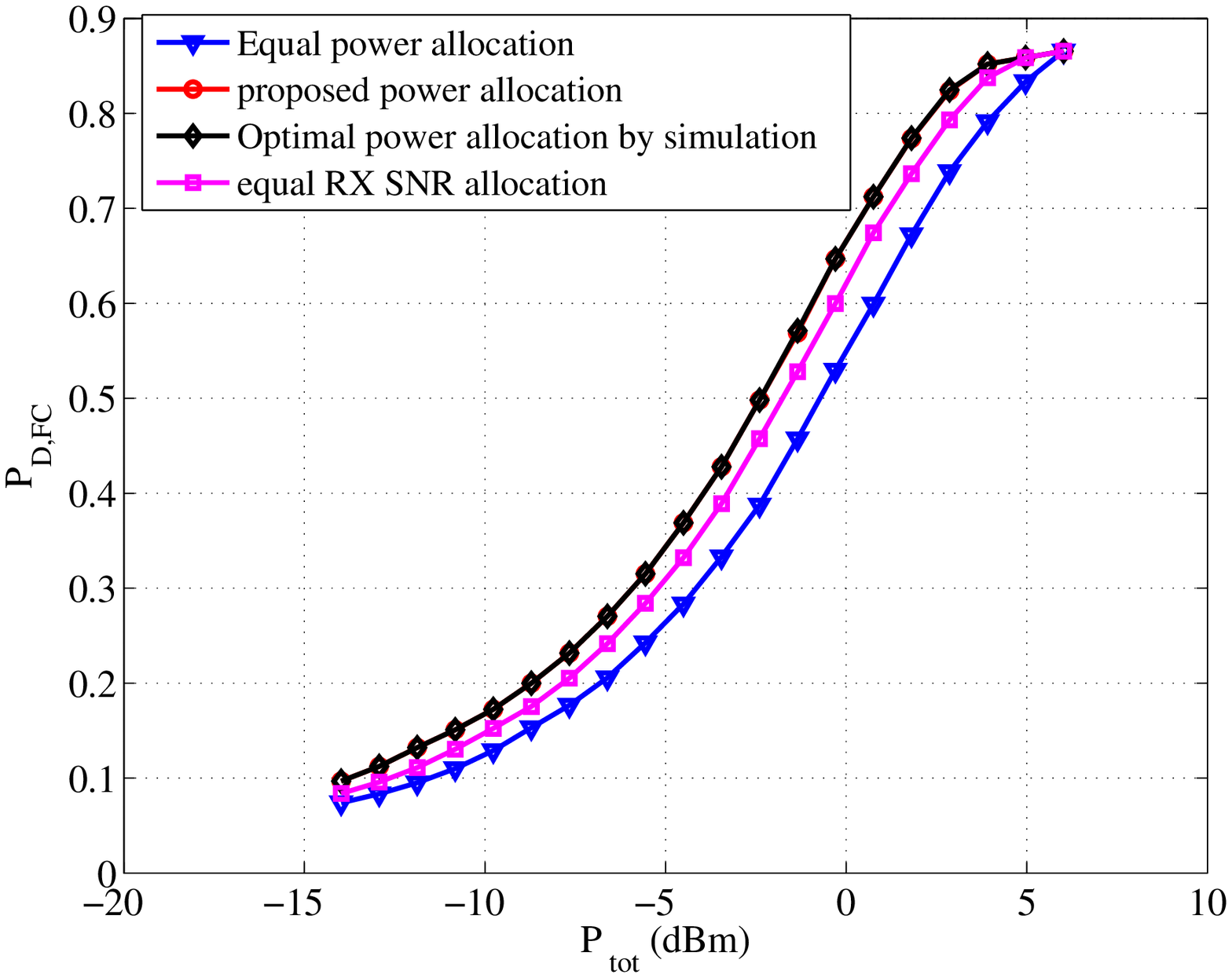}
\caption{\label{fig:two sensor nonorth channel PDFC vs Ptot case
4} The FC detection probability $P_{D,FC}$ as a function of
$P_{\rm tot}$ of Case \ref{subsec:two sensors with non-orthogonal
channels}4.}
\end{minipage}
\hfill
\end{figure*}

\subsection{Ten sensors with orthogonal MIMO channels}
\label{subsec:ten sensors with orthogonal channels}

Here we consider ten sensors scattered around an FC. We will
investigate five cases, according to various sensor distance and
detection probability combinations. In Case \ref{subsec:ten
sensors with orthogonal channels}1, Case \ref{subsec:ten sensors
with orthogonal channels}2, and Case \ref{subsec:ten sensors with
orthogonal channels}4, some of the sensors do not operate in
region $\mathbf{\cal S}$, so the interior point optimization
algorithm is used in these three cases. In Case \ref{subsec:ten
sensors with orthogonal channels}3 and Case \ref{subsec:ten
sensors with orthogonal channels}5, all the sensors operate in
region $\mathbf{\cal S}$, thus Algorithm 1 is used. The total
power budget $P_{\rm tot}$ varies from -7 dBm to 13 dBm (when each
sensor transmits at full power 2 mW).

\subsubsection{Case \ref{subsec:ten sensors with orthogonal channels}1}
The distance between sensor $j$ and the FC is $d_j=2+0.6(j-1)$ m,
e.g. $d_1=2$ m and $d_{10}=7.4$ m. Sensor $j$ has detection
probability of $P_D(j)=0.1+0.09(j-1)$, e.g. $P_D(1)=0.1$ and
$P_D(10)=0.91$. In this case, sensors closer to the FC have worse
detection probability.

\begin{table}[tb]
\begin{tabular}{|m{0.08\linewidth}||m{0.03\linewidth}|m{0.03\linewidth}|m{0.03\linewidth}|m{0.03\linewidth}
|m{0.03\linewidth}|m{0.03\linewidth}|m{0.03\linewidth}|m{0.03\linewidth}|m{0.03\linewidth}|m{0.03\linewidth}|}\hline
 $P_{\rm tot}$ (dBm)& 1 & 2 & 3 & 4 & 5 & 6 & 7 & 8 & 9 & 10 \\ \hline \hline
 -7.0 & 0 & 0 & 0 & 0 & 0 & 7 & 15 & 21 & 26 & 31 \\ \hline
 -2.8 & 0 & 0 & 0 & 0 & 6 & 11 & 16 & 19 & 23 & 25 \\ \hline
 3.5 & 0 & 0 & 0 & 5 & 9 & 12 & 15 & 17 & 20 & 22 \\ \hline
 8.8 & 0 & 3 & 7 & 9 & 10 & 11 & 13 & 14 & 16 & 17 \\ \hline
 13 & 10 & 10 & 10 & 10 & 10 & 10 & 10 & 10 & 10 & 10 \\ \hline
\end{tabular}
\caption{\label{table:10SensorCase1} Percentage of the total power
allocated to each sensor for for case \ref{subsec:ten sensors with
orthogonal channels}1.}
\end{table}

\begin{table}[tb]
\begin{tabular}{|m{0.08\linewidth}||m{0.03\linewidth}|m{0.03\linewidth}|m{0.03\linewidth}|m{0.03\linewidth}
|m{0.03\linewidth}|m{0.03\linewidth}|m{0.03\linewidth}|m{0.03\linewidth}|m{0.03\linewidth}|m{0.03\linewidth}|}\hline
 $P_{\rm tot}$ (dBm)& 1 & 2 & 3 & 4 & 5 & 6 & 7 & 8 & 9 & 10 \\ \hline \hline
 -7.0 & 100 & 0 & 0 & 0 & 0 & 0 & 0 & 0 & 0 & 0 \\ \hline
 -2.8 & 74 & 26 & 0 & 0 & 0 & 0 & 0 & 0 & 0 & 0  \\ \hline
 3.5 & 36 & 23 & 15 & 10 & 7 & 5 & 3 & 1 & 0 & 0 \\ \hline
 8.8 & 18 & 14 & 12 & 10 & 9 & 8 & 8 & 7 & 7 & 7 \\ \hline
 13 & 10 & 10 & 10 & 10 & 10 & 10 & 10 & 10 & 10 & 10 \\ \hline
\end{tabular}
\caption{\label{table:10SensorCase2} Percentage of total power
allocated to each sensor for case \ref{subsec:ten sensors with
orthogonal channels}2.}
\end{table}

The percentage of the total power budget $P_{\rm tot}$ allocated
to each sensor is shown in Table \ref{table:10SensorCase1}. When
$P_{\rm tot}$ is low, more power is distributed to sensors farther
away from the FC. Intuitively this is because even though sensors
closer to the FC have good channel gain, their local detection
qualities are much worse than those of the farther sensors. With
$P_{\rm tot}$ increases, power is distributed more evenly among
the sensors, because some sensors have already reached their
maximum output power. Eventually, when the total power budget
reaches 13 dBm, every sensor transmits at $P_{\rm max}=3$ dBm.

\begin{figure*}[tb]
\begin{minipage}{0.32\linewidth}
\includegraphics[width=\textwidth]{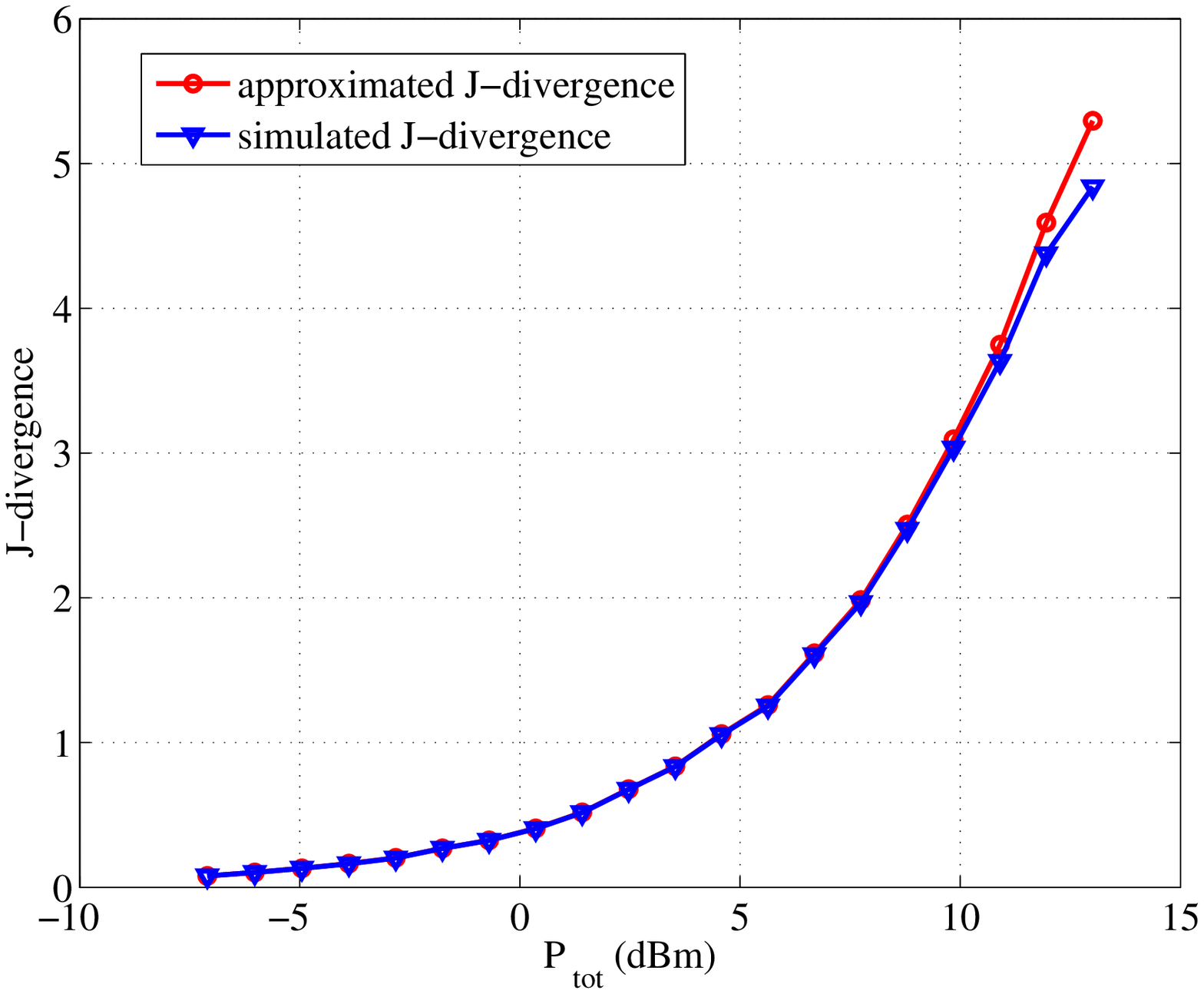}
\caption{\label{fig:J_Ptot_10sensor_case1} Approximated and
simulated J-divergence as a function of $P_{\rm tot}$ for the
proposed power allocation of case \ref{subsec:ten sensors with
orthogonal channels}1.}
\end{minipage}
\hfill
\begin{minipage}{0.32\linewidth}
\includegraphics[width=\textwidth]{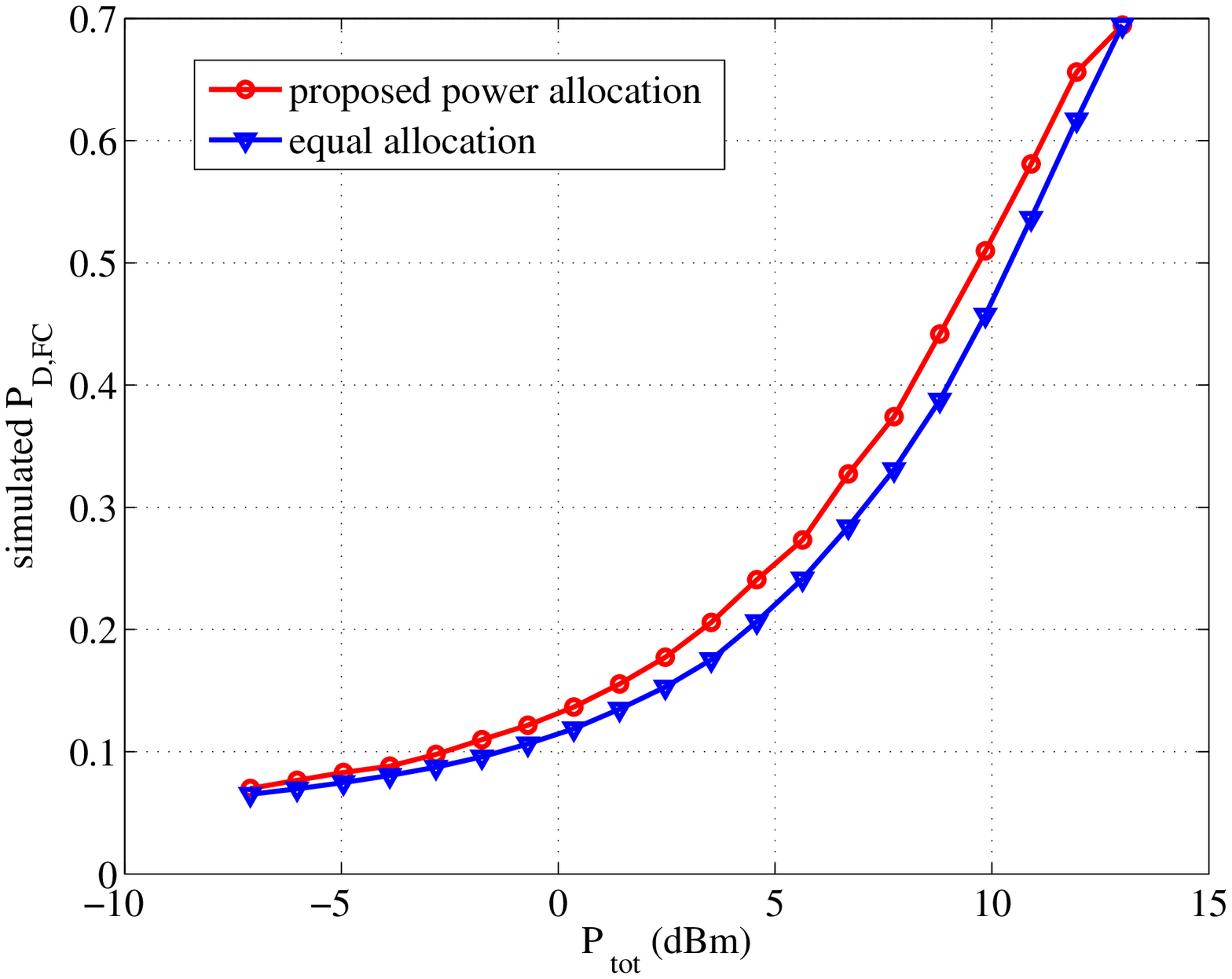}
\caption{\label{fig:PD_Ptot_10sensor_case1} Simulated $P_{D,FC}$
as a function of $P_{\rm tot}$ for the proposed power allocation
and equal power allocation of case \ref{subsec:ten sensors with
orthogonal channels}1.}
\end{minipage}
\hfill
\begin{minipage}{0.32\linewidth}
\includegraphics[width=\textwidth]{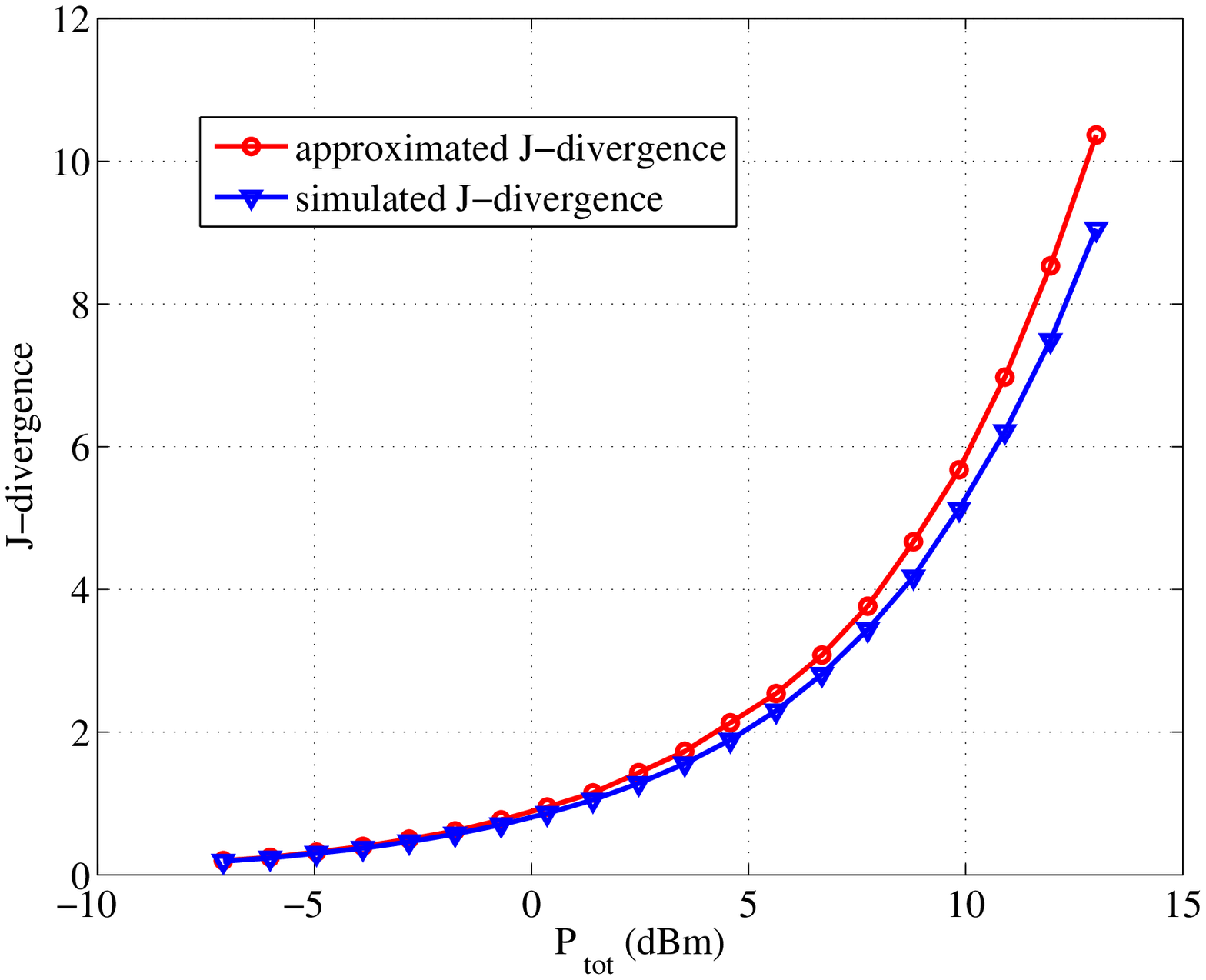}
\caption{\label{fig:J_Ptot_10sensor_case2} Approximated and
simulated J-divergence as a function of $P_{\rm tot}$ for the
proposed power allocation of case \ref{subsec:ten sensors with
orthogonal channels}2.}
\end{minipage}
\hfill
\end{figure*}

Figure \ref{fig:J_Ptot_10sensor_case1} shows that the approximated
and actual (by Monte Carlo simulation) J-divergences are very
close even when every sensor is transmitting at full power. The
maximum received SNR of the closest sensor at the FC is about 12
dB. Similar to Section \ref{subsec:two sensors with orthogonal
channels}, we can see that the approximation in J-divergence works
well with SNR as high as 12 dB.

Figure \ref{fig:PD_Ptot_10sensor_case1} shows the simulated
detection probability at the FC $P_{D,FC}$ of the proposed power
allocation and the equal power allocation\footnote{For this
scenario with 10 sensors, the complexity is too high to find the
optimal power allocation by brute-force search and simulations in
a ten-dimensional space of all possible power allocations. So in
this scenario, we will not include the optimal power allocation
found by simulation that gives the highest $P_{D,FC}$.}. In this
case, the proposed power allocation can use about 1 dB less power
than equal power allocation to reach the same detection
performance at the FC.

\subsubsection{Case \ref{subsec:ten sensors with orthogonal channels}2}
The distance between sensor $j$ and the FC is $d_j=2+0.6(j-1)$ m.
Sensor $j$ has detection probability of $P_D(j)=0.4+0.06(j-1)$,
e.g. $P_D(1)=0.4$ and $P_D(10)=0.94$. In this case, sensors closer
to the FC still have worse detection probability, but the gap in
local detection qualities is not as large as that in case 1.

In this case, Table \ref{table:10SensorCase2} shows that when
$P_{\rm tot}$ is low, more power is distributed to sensors closer
to the FC. The advantage in channel gain of sensors closer to the
FC has offset their disadvantage in detection quality. The water
filling effect is obvious here, and the sensors get positive power
allocation in a sequential fashion.

In Figure \ref{fig:J_Ptot_10sensor_case2}, the approximated and
actual J-divergences are still very close. Figure
\ref{fig:PD_Ptot_10sensor_case2} shows the proposed power
allocation has a maximum power saving of 4 dB (more than 50\%)
compared to equal power allocation.

\subsubsection{Case \ref{subsec:ten sensors with orthogonal channels}3}
$d_j=2+0.6(j-1)$ m, and $P_D(j)=0.8$.
\begin{table}[tb]
\begin{tabular}{|m{0.08\linewidth}||m{0.03\linewidth}|m{0.03\linewidth}|m{0.03\linewidth}|m{0.03\linewidth}
|m{0.03\linewidth}|m{0.03\linewidth}|m{0.03\linewidth}|m{0.03\linewidth}|m{0.03\linewidth}|m{0.03\linewidth}|}\hline
 $P_{\rm tot}$ (dBm)& 1 & 2 & 3 & 4 & 5 & 6 & 7 & 8 & 9 & 10 \\ \hline \hline
 -7.0 & 100 & 0 & 0 & 0 & 0 & 0 & 0 & 0 & 0 & 0 \\ \hline
 -2.8 & 81 & 19 & 0 & 0 & 0 & 0 & 0 & 0 & 0 & 0  \\ \hline
 3.5 & 54 & 33 & 13 & 0 & 0 & 0 & 0 & 0 & 0 & 0 \\ \hline
 8.8 & 26 & 26 & 21 & 15 & 8 & 3 & 0 & 0 & 0 & 0 \\ \hline
 13 & 10 & 10 & 10 & 10 & 10 & 10 & 10 & 10 & 10 & 10 \\ \hline
\end{tabular}
\caption{\label{table:10SensorCase3} Percentage of total power
allocated to each sensor for case \ref{subsec:ten sensors with
orthogonal channels}3.}
\end{table}

\begin{table}[tb]
\begin{tabular}{|m{0.08\linewidth}||m{0.03\linewidth}|m{0.03\linewidth}|m{0.03\linewidth}|m{0.03\linewidth}
|m{0.03\linewidth}|m{0.03\linewidth}|m{0.03\linewidth}|m{0.03\linewidth}|m{0.03\linewidth}|m{0.03\linewidth}|}\hline
 $P_{\rm tot}$ (dBm)& 1 & 2 & 3 & 4 & 5 & 6 & 7 & 8 & 9 & 10 \\ \hline \hline
 -7.0 & 100 & 0 & 0 & 0 & 0 & 0 & 0 & 0 & 0 & 0 \\ \hline
 -2.8 & 100 & 0 & 0 & 0 & 0 & 0 & 0 & 0 & 0 & 0  \\ \hline
 3.5 & 73 & 27 & 0 & 0 & 0 & 0 & 0 & 0 & 0 & 0 \\ \hline
 8.8 & 26 & 26 & 26 & 15 & 7 & 0 & 0 & 0 & 0 & 0 \\ \hline
 13 & 10 & 10 & 10 & 10 & 10 & 10 & 10 & 10 & 10 & 10 \\ \hline
\end{tabular}
\caption{\label{table:10SensorCase4} Percentage of total power
allocated to each sensor for case \ref{subsec:ten sensors with
orthogonal channels}4.}
\end{table}
In Table \ref{table:10SensorCase3} power allocation is even more
biased toward sensors closer to the FC compared to Case
\ref{subsec:ten sensors with orthogonal channels}2. All the
sensors now have the same detection quality, but the sensors
closer to the FC have the advantage in channel gain.

\begin{figure*}[tb]
\begin{minipage}{0.32\linewidth}
\includegraphics[width=\textwidth]{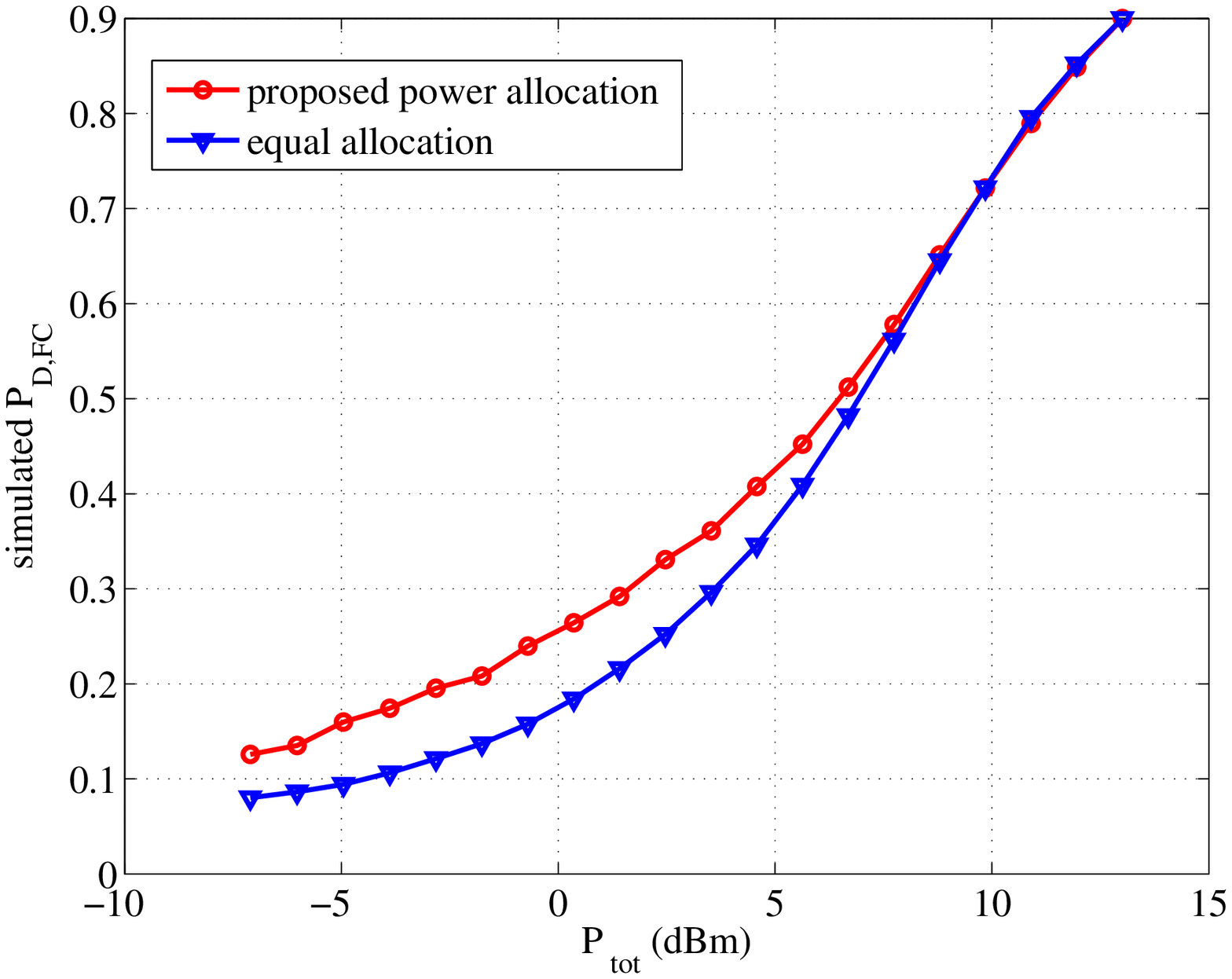}
\caption{\label{fig:PD_Ptot_10sensor_case2} Simulated $P_{D,FC}$
as a function of $P_{\rm tot}$ for the proposed power allocation
and equal power allocation of case \ref{subsec:ten sensors with
orthogonal channels}2.}
\end{minipage}
\hfill
\begin{minipage}{0.32\linewidth}
\includegraphics[width=\textwidth]{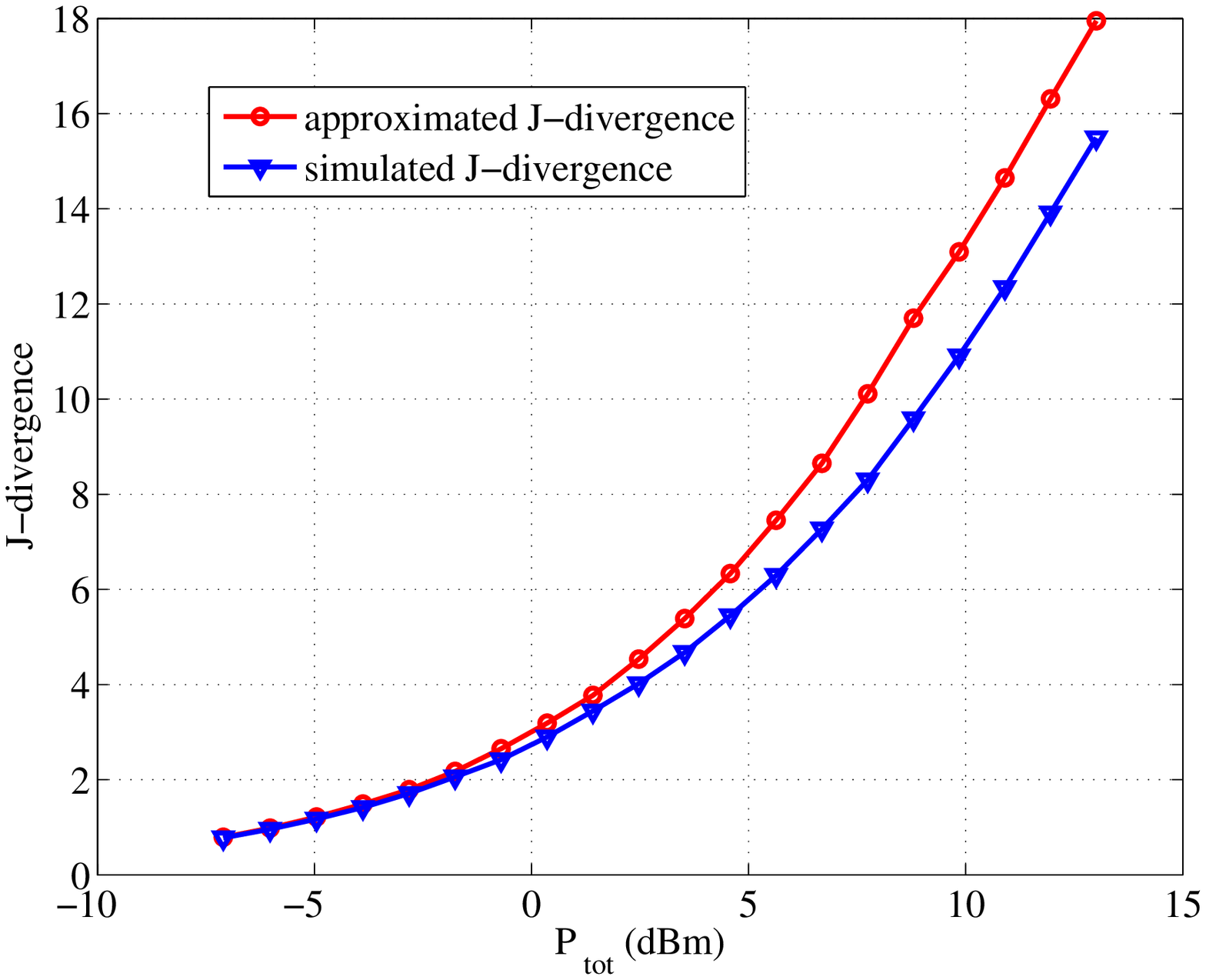}
\caption{\label{fig:J_Ptot_10sensor_case3} Approximated and
simulated J-divergence as a function of $P_{\rm tot}$ for the
proposed power allocation of case \ref{subsec:ten sensors with
orthogonal channels}3.}
\end{minipage}
\hfill
\begin{minipage}{0.32\linewidth}
\includegraphics[width=\textwidth]{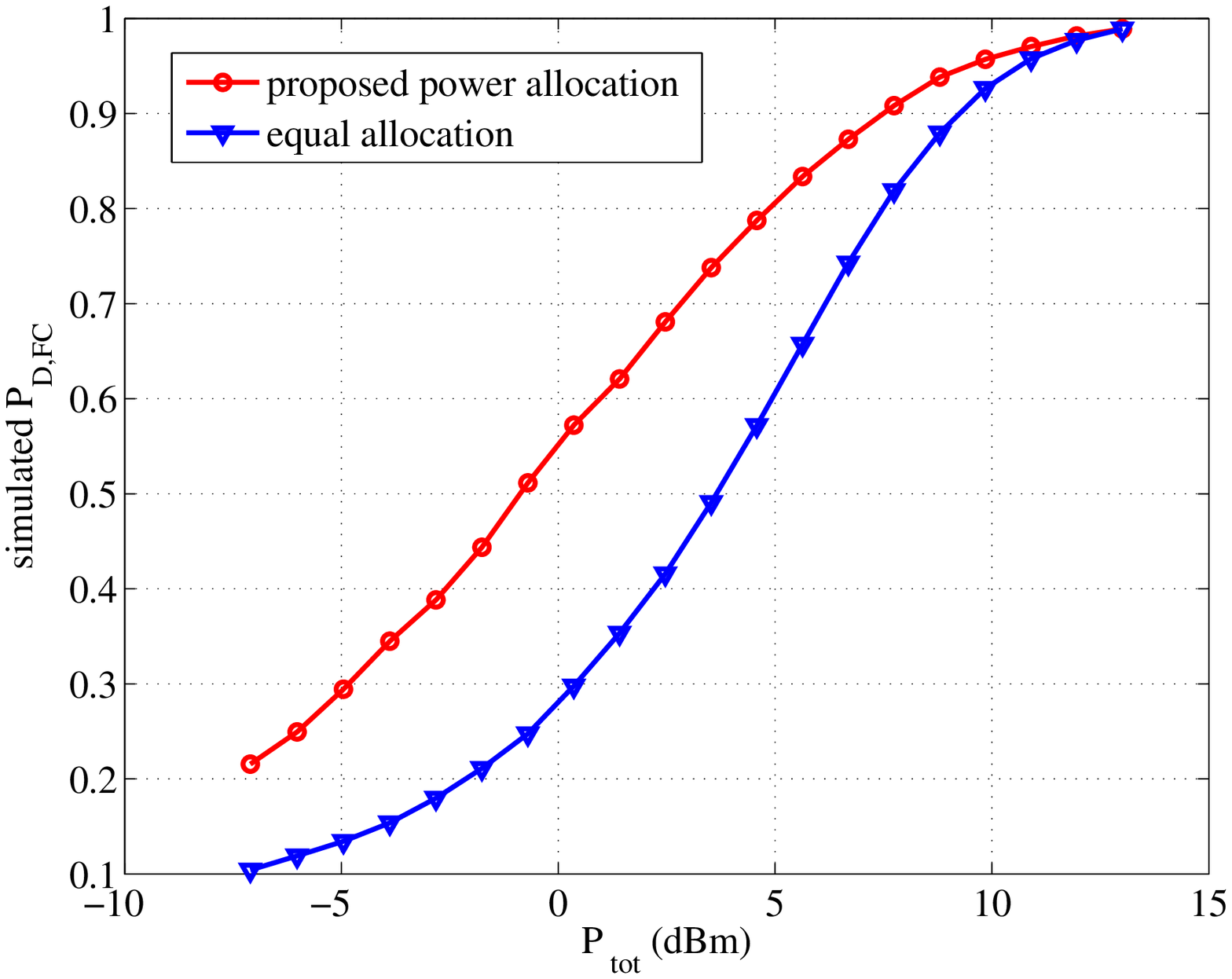}
\caption{\label{fig:PD_Ptot_10sensor_case3} Simulated $P_{D,FC}$
as a function of $P_{\rm tot}$ for the proposed power allocation
and equal power allocation of case \ref{subsec:ten sensors with
orthogonal channels}3.}
\end{minipage}
\hfill
\end{figure*}

In Figure \ref{fig:J_Ptot_10sensor_case3}, the approximated and
actual J-divergences start to show some difference. Figure
\ref{fig:PD_Ptot_10sensor_case3} shows the proposed power
allocation has a maximum power saving of more than 5 dB.

\subsubsection{Case \ref{subsec:ten sensors with orthogonal channels}4}
$d_j=2+0.6(j-1)$ m, and $P_D(j)=0.94-0.06(j-1)$. In this case, the
sensors closer to the FC have advantage in both channel gain and
local detection quality. So Table \ref{table:10SensorCase4} power
allocation is even more biased toward sensors closer to the FC
compared with Case \ref{subsec:ten sensors with orthogonal
channels}3.

In Figure \ref{fig:J_Ptot_10sensor_case4}, the approximated and
actual J-divergences have more difference than the previous three
cases, but their shapes are still quite similar. Figure
\ref{fig:PD_Ptot_10sensor_case4} shows that the proposed power
allocation consumes only less than 25\% (has more than 6 dB
saving) of the total power used by the equal power allocation to
achieve the same detection performance at the FC.

\subsubsection{Case \ref{subsec:ten sensors with orthogonal channels}5}
\begin{figure*}[tb]
\begin{minipage}{0.32\linewidth}
\includegraphics[width=\textwidth]{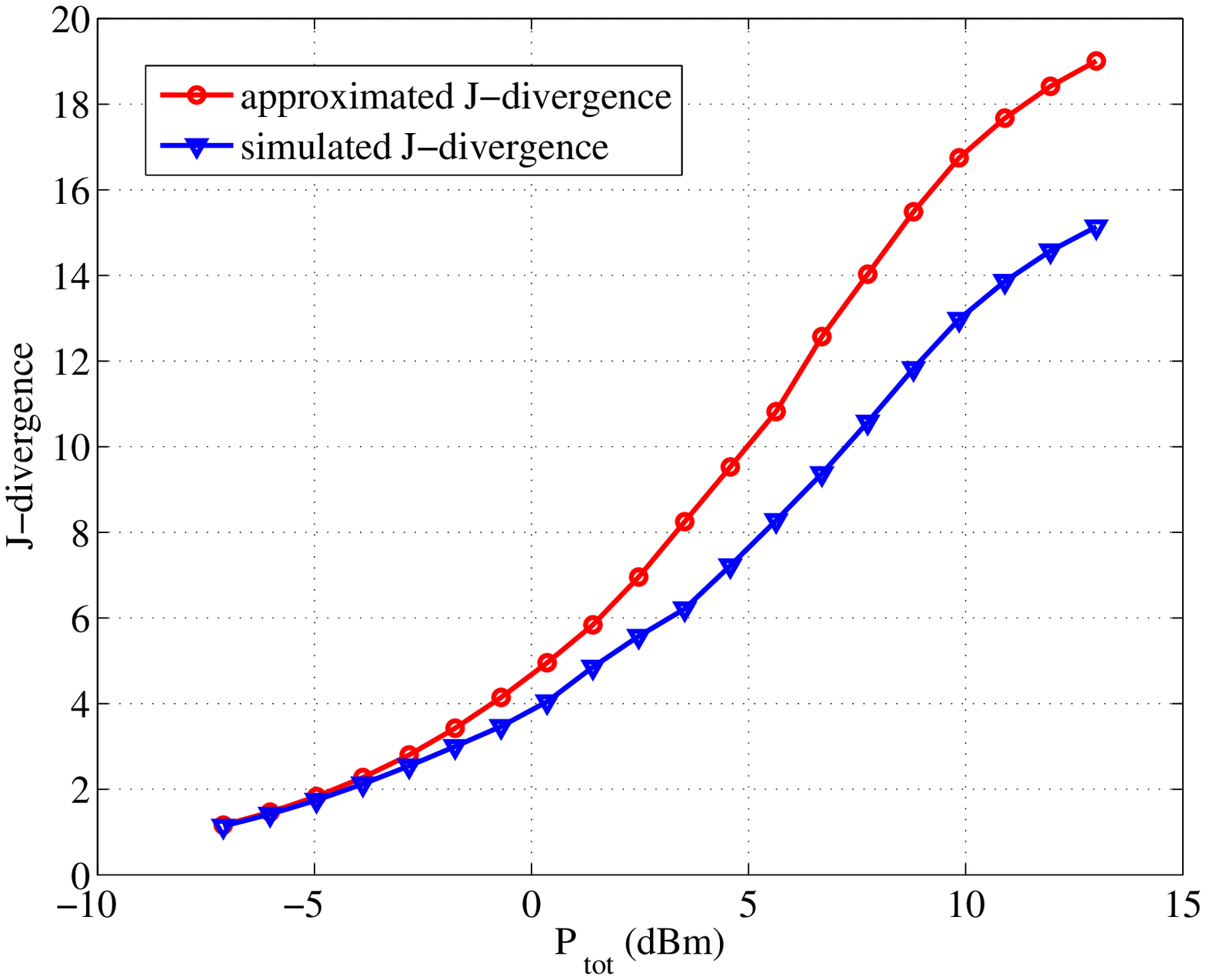}
\caption{\label{fig:J_Ptot_10sensor_case4} Approximated and
simulated J-divergence as a function of $P_{\rm tot}$ for the
proposed power allocation of case \ref{subsec:ten sensors with
orthogonal channels}4.}
\end{minipage}
\hfill
\begin{minipage}{0.32\linewidth}
\includegraphics[width=\textwidth]{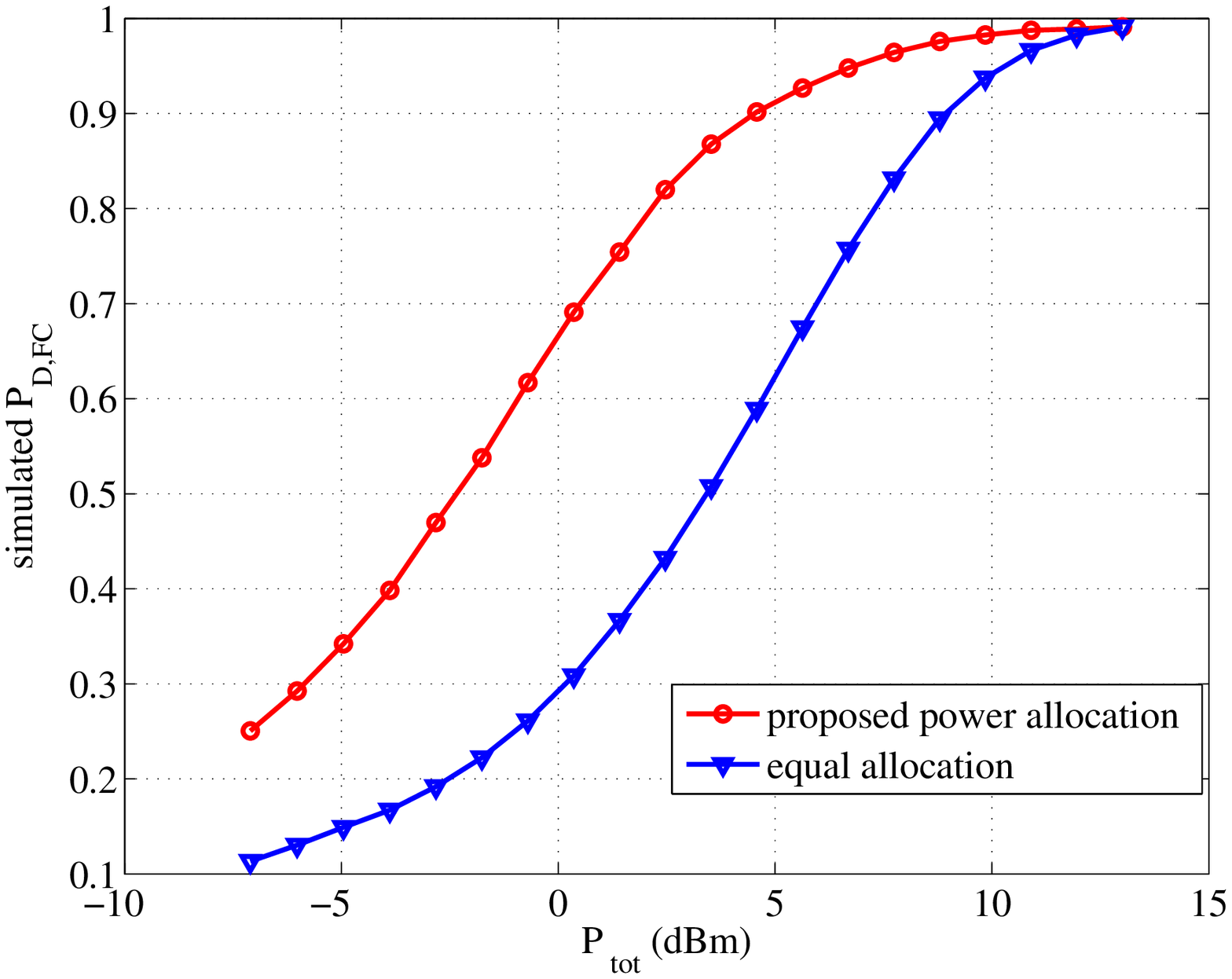}
\caption{\label{fig:PD_Ptot_10sensor_case4} Simulated $P_{D,FC}$
as a function of $P_{\rm tot}$ for the proposed power allocation
and equal power allocation of case \ref{subsec:ten sensors with
orthogonal channels}4.}
\end{minipage}
\hfill
\begin{minipage}{0.32\linewidth}
\includegraphics[width=\textwidth]{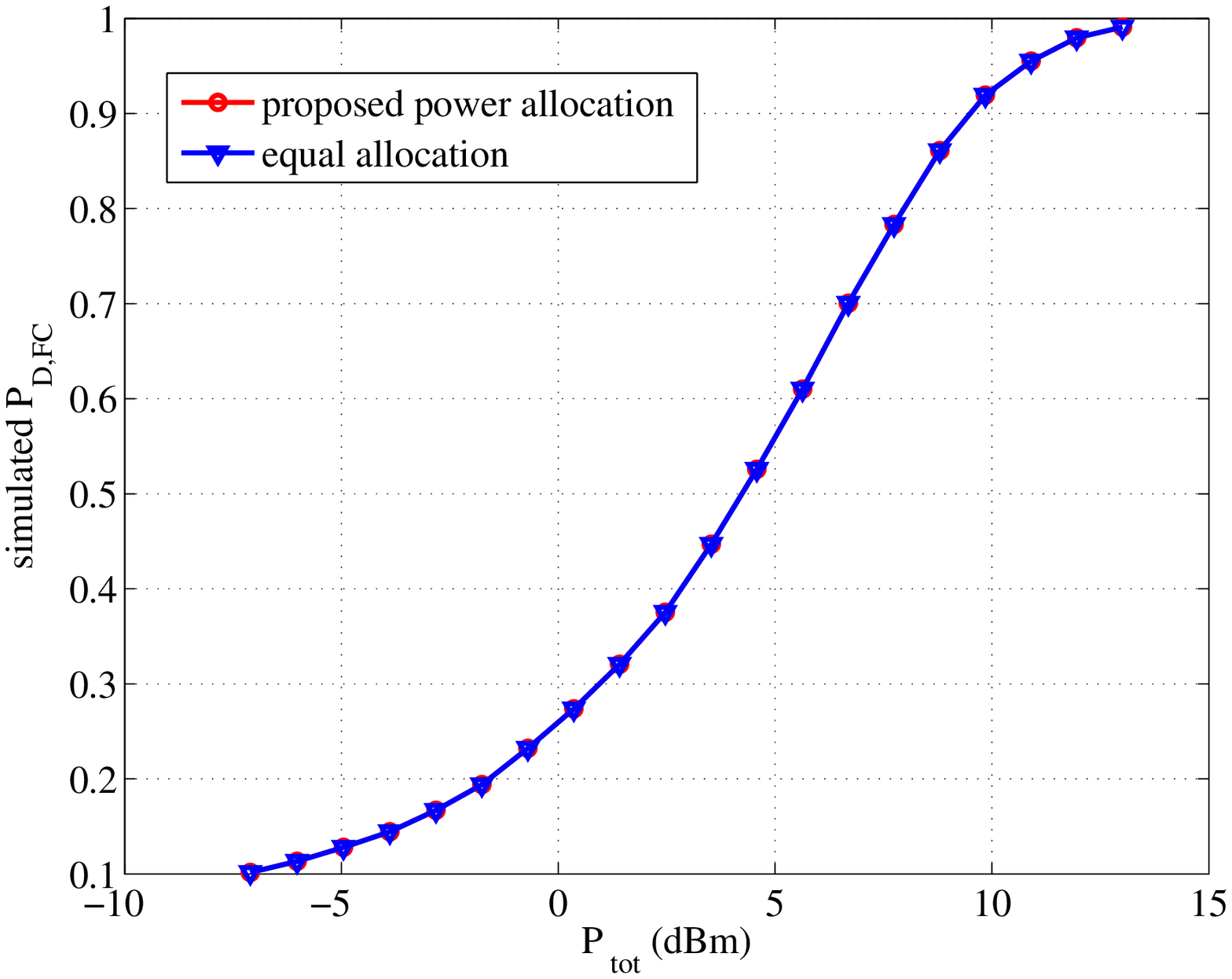}
\caption{\label{fig:PD_Ptot_10sensor_case5} Simulated $P_{D,FC}$
as a function of $P_{\rm tot}$ for the proposed power allocation
and equal power allocation of case \ref{subsec:ten sensors with
orthogonal channels}5. (Note that the curves for $\circ$ and
$\triangledown$ overlay in this graph.)}
\end{minipage}
\hfill
\end{figure*}

$d_j=4$ m, and $P_D(j)=0.8$. In this case, all sensors have the
same detection probability and distance from the FC. This case
serves as a sanity check because intuitively the sensors should
always have equal power allocation in this case. Table
\ref{table:10SensorCase5} and Figure
\ref{fig:PD_Ptot_10sensor_case5} verify this intuition.

\begin{table}[tb]
\centering
\begin{tabular}{|m{0.08\linewidth}||m{0.03\linewidth}|m{0.03\linewidth}|m{0.03\linewidth}|m{0.03\linewidth}
|m{0.03\linewidth}|m{0.03\linewidth}|m{0.03\linewidth}|m{0.03\linewidth}|m{0.03\linewidth}|m{0.03\linewidth}|}\hline
 $P_{\rm tot}$ (dBm)& 1 & 2 & 3 & 4 & 5 & 6 & 7 & 8 & 9 & 10 \\ \hline \hline
 -7.0 & 10 & 10 & 10 & 10 & 10 & 10 & 10 & 10 & 10 & 10 \\ \hline
 -2.8 & 10 & 10 & 10 & 10 & 10 & 10 & 10 & 10 & 10 & 10  \\ \hline
 3.5 & 10 & 10 & 10 & 10 & 10 & 10 & 10 & 10 & 10 & 10 \\ \hline
 8.8 & 10 & 10 & 10 & 10 & 10 & 10 & 10 & 10 & 10 & 10 \\ \hline
 13 & 10 & 10 & 10 & 10 & 10 & 10 & 10 & 10 & 10 & 10 \\ \hline
\end{tabular}
\caption{\label{table:10SensorCase5} Percentage of total power
allocated to each sensor for case \ref{subsec:ten sensors with
orthogonal channels}5.}
\end {table}

\section{Conclusions}
\label{sec:conclusions} In this paper we have studied the problem
of optimal power allocation for distributed detection over MIMO
channels in wireless sensor networks. Our contribution is novel
compared to the pervious work in the following senses: (1) We have
considered a distributed detection system with a MIMO channel to
account for non-ideal communications between the sensors and the
FC. (2) We have assumed that there are a finite number of sensors
and the sensors have independent but nonidentically distributed
observations. (3) We also have assumed both individual and joint
constraints on the power that the sensors can expend to transmit
their local decisions to the FC. (4) We have developed a power
allocation scheme to distribute the total power budget among the
sensors so that the detection performance at the FC is optimized
in terms of the J-divergence. (5) The proposed power allocation is
a tradeoff between the quality of the local decisions of the
sensors and the quality of the communication channels between the
sensors and the FC. Simulations show that, to achieve the same
detection performance at the FC, the proposed power allocation can
use as little as 25\% of the total power used by equal power
allocation.

\appendices

\section{Proof of Lemma \ref{lem: special case J non decreasing}}
\label{proof:lem 4}
\begin{proof}
Taking the derivative of $J(P_1,\cdots,P_K)$ in \eqref{eqn:J
approx optimization special case} with respect to $P_j$, we have
\begin{align}
&\frac{\partial}{\partial P_j} J(P_1,\cdots,P_K)\Big
|_{\{P_i\}_{i=1}^K, \; P_i\ge0} \nonumber \\
&=\frac{(\alpha_F(j)-\beta_F(j))\sigma^2g_j}
{(\sigma^2+\beta_F(j)g_jP_j)^2}+\frac{(\alpha_D(j)-\beta_D(j))\sigma^2g_j}{(\sigma^2+\beta_D(j)g_jP_j)^2}
\nonumber \\
&=c\big[(\alpha_F(j)-\beta_F(j))(\sigma^2+\beta_D(j)g_jP_j)^2
\notag \\ & \qquad +
(\alpha_D(j)-\beta_D(j))(\sigma^2+\beta_F(j)g_jP_j)^2\big],
\nonumber \\
&=c\big[d_0\sigma^4+2d_1\sigma^2g_jP_j+d_2g_j^2P_j^2\big],
\end{align} where
\begin{align}
c&=\frac{\sigma^2g_j}
{(\sigma^2+\beta_F(j)g_jP_j)^2(\sigma^2+\beta_D(j)g_jP_j)^2}, \\
d_0&=\alpha_F(j)+\alpha_D(j)-\beta_F(j)-\beta_D(j), \\
d_1&=\alpha_F(j)\beta_D(j)+\alpha_D(j)\beta_F(j)-2\beta_F(j)\beta_D(j), \\[-3mm]
\intertext{and}\notag \\[-8mm]
d_2&=\alpha_F(j)\beta_D(j)^2+\alpha_D(j)\beta_F(j)^2 \notag \\
&\qquad -\beta_F(j)^2\beta_D(j)-\beta_F(j)\beta_D(j)^2.
\end{align}

Substituting $\alpha_F(j)$, $\beta_F(j)$, $\alpha_D(j)$,
$\beta_D(j)$ into equations \eqref{eqn:alpha_D beta_D} --
\eqref{eqn:alpha_D beta_D end}, and after lengthy algebra, we have
\begin{align}
d_0&=2(P_D(j)-P_F(j))^2\ge 0, \\
d_1&=(P_D(j)-P_F(j))^2\Big[P_D(j)P_F(j) \notag
\\ &\qquad+(1-P_D(j))(1-P_F(j))\Big]\ge 0,
\\[-3mm]
\intertext{and}\notag \\[-8mm]
d_2&=(P_D(j)-P_F(j))^2\Big[(P_D(j)+P_F(j)) \notag \\ & \qquad (P_D(j)+P_F(j)-1-P_D(j)P_F(j))^2 \nonumber \\
& \quad \quad+P_D(j)^2P_F(j)^2(2-P_D(j)-P_F(j))\Big]\ge0.
\end{align}
Since $c$ and $P_j$ are also nonnegative, we conclude that
\begin{align}
&\frac{\partial}{\partial P_j} J(P_1,\cdots,P_K)\Big
|_{\{P_i\}_{i=1}^K, \; P_i\ge0} \ge 0.
\end{align}
\end{proof}

\section{Proof of Lemma \ref{lem: special case J second derivative}}
\label{proof:lem 5}
\begin{proof}
From \eqref{eqn:2nd derivative of J(P)}, we can easily see that
$\frac{\partial^2}{\partial P_j^2}J(P_1,\cdots,P_K)\le 0$ if and
only if $C_i\ge0$, $i=0,\cdots,3$. From the proof of Lemma
\ref{lem: special case J non decreasing}, we know that $C_i\ge0$,
$i=1,2,3$, so $\frac{\partial^2}{\partial
P_j^2}J(P_1,\cdots,P_K)\le 0$ if and only if $C_0\ge0$.

We have
\begin{align}
C_0&=\beta_F(j)\Big[\alpha_F(j)-\beta_F(j)\Big]+\beta_D(j)\Big[\alpha_D(j)-\beta_D(j)\Big].
\end{align} Substituting $\alpha_F(j)$, $\beta_F(j)$, $\alpha_F(j)$, and $\beta_F(j)$
into \eqref{eqn:alpha_D beta_D} -- \eqref{eqn:alpha_D beta_D end}
and after some algebra, we have
\begin{align}
C_0&=\Big(P_D(j)-P_F(j)\Big)^2\Big[-2P_D(j)^2 \notag
\\ & \qquad+(3-2P_F(j))P_D(j) +3P_F(j)-2P_F(j)^2-1\Big].
\end{align}
Now, $C_0\ge0$ if and only if
\begin{align}
-2P_D(j)^2 &+(3-2P_F(j))P_D(j) \notag \\
& \qquad +3P_F(j)-2P_F(j)^2-1\ge0.
\end{align}
Solving the above quadratic inequality leads to Lemma \ref{lem:
special case J second derivative}

\end{proof}


\begin{thebibliography}{99}
\addtolength{\baselineskip}{-.2\baselineskip}

\bibitem{Ali Silvey 1966}
S.\ M.\ Ali and S.\ D.\ Silvey, ``A general class of coefficients
of divergence of one distribution from another,'' {\em J.\ Royal
Stat. Soc.}, Series B, vol. 28, pp. 131-142, 1966.

\bibitem{Bertsekas99book}
D.\ P.\ Bertsekas, {\em Nonlinear Programming}, $2^{nd}$ ed.,
Athena Scientific, Boston, 1999.

\bibitem{Blum Kassam Poor 1997}
R.\ S.\ Blum, S.\ A.\ Kassam, and H.\ V.\ Poor, ``Distributed
detection with multiple sensors: Part II -- Advanced topics,''
{\em Proc. IEEE}, vol. 85, no. 1, pp. 64-79, Jan. 1997.

\bibitem{Boyd_book}
S.\ Boyd and L.\ Vandenberghe, {\em Convex Optimization}, John
Wiley \& Sons, Inc., New York, 1991.

\bibitem{Chen Jiang Kasetkasem Varshney T-SP 2004}
B.\ Chen, R.\ Jiang, T.\ Kasetkasem, and P.\ K.\ Varshney,
``Channel aware decision fusion in wireless sensor networks,''
{\em IEEE Trans. Signal Processing}, vol. 52, no. 12, pp.
3454–3458, Dec. 2004.

\bibitem{Chen Willett T-IT 2005}
B.\ Chen and P.\ K.\ Willett, ``On the optimality of the
likelihood-ratio test for local sensor decision rules in the
presence of nonideal channels,'' {\em IEEE Trans. Inform. Theory},
vol. 51, no. 2, pp. 693–699, Feb. 2005.

\bibitem{Chamberland Veeravalli 2003}
J.\ F.\ Chamberland and V.\ V.\ Veeravalli, ``Decentralized
detection in sensor networks,'' {\em IEEE Trans. Signal
Processing}, vol. 51, no. 2, pp. 407-416, Feb. 2003.

\bibitem{Chamberland Veeravalli 2004}
J.\ F.\ Chamberland and V.\ V.\ Veeravalli, ``Asymptotic results
for decentralized detection in power constrained wireless sensor
networks,'' {\em IEEE Journal on Selected Areas in
Communications}, vol. 22, no. 6, pp. 1007-1015, Aug. 2004.

\bibitem{Chamberland Veeravalli 2006}
J.\ F.\ Chamberland and V.\ V.\ Veeravalli, ``How dense should a
sensor network be for detection with correlated observations?''
{\em IEEE Trans. Inform. Theory}, vol. 52, no. 11, pp. 5099–5106,
Nov. 2006.

\bibitem{Chamberland Veeravalli 2007}
J.\ F.\ Chamberland and V.\ V.\ Veeravalli, ``Wireless sensors in
distributed detection applications,'' {\em IEEE Signal Processing
Magazine}, vol. 24, no. 3, pp. 16-25, May 2007.

\bibitem{Cover_book}
T.\ Cover and J.\ Thomas, {\em Elements of Information Theory},
John Wiley \& Sons, Inc., New York, 1991.

\bibitem{Csiszar_book}
I.\ Csisz\'{a}r and J.\ K\"{o}rner, {\em Information Theory:
Coding Theorems for Discrete Memoryless Systems}, Akad\'{e}miai
Kaid\'{o}, Budapest, 1981.

\bibitem{Duman Salehi 1998}
T.\ M.\ Duman and M.\ Salehi, ``Decentralized detection over
multiple-access channels,'' {\em IEEE Trans. Aerosp. Electron.
Syst.}, vol. 34, no. 2, pp. 469-476, Apr. 1998.

\bibitem{Golub_book}
G.\ H.\ Golub and C.\ F.\ Van Loan, {\em Matrix Computations},
Johns Hopkins University Press, Baltimore, MD, 1996.

\bibitem{Jayaweera T-SP 2007}
S. K. Jayaweera, ``Bayesian fusion performance and system
optimization for distributed stochastic Gaussian signal detection
under communication constraints,'' {\em IEEE Trans. Signal
Processing}, vol. 55, no. 4, pp. 1238-1250, Apr. 2007.

\bibitem{Jeffreys 1946}
H.\ Jeffreys, ``An invariant form for the prior probability in
estimation problems,'' {\em Proc. Roy. Soc. A.}, vol. 186, pp.
453-461, 1946.

\bibitem{Kailath 1967}
T.\ Kailath, ``The divergence and Bhattacharyya distance measures
in signal selection,'' {\em IEEE Trans. Communication Technology},
vol. 15, No. 2, pp. 52-60, Feb. 1967.

\bibitem{Kobayashi Thomas 1967}
H.\ Kobayashi and J.\ B.\ Thomas, ``Distance measures and related
criteria,'' {\em Proc. Fifth Annual Allerton Conf. Circuit and
System Theory}, pp. 491-500, Oct. 1967.

\bibitem{Kobayashi 1970}
H.\ Kobayashi, ``Distance measures and asymptotic relative
efficiency,'' {\em IEEE Trans. Inform. Theory}, vol. 16, No. 3,
pp. 288-291, May 1970.

\bibitem{Liu Sayeed T-SP 2007}
K. Liu and A.M. Sayeed, ``Type-based decentralized detection in
wireless sensor networks,'' {\em IEEE Trans. Signal Processing},
vol. 55, no. 5, pp. 1899-1910, May 2007.

%
\bibitem{Mergen Naware Tong T-SP 2007}
G.\ Mergen, V.\ Naware, and L.\ Tong, ``Asymptotic detection
performance of type-based multiple access over multiaccess fading
channels,'' {\em IEEE Trans. Signal Processing}, vol. 55, no. 3,
pp. 1081-1092, Mar. 2007.

\bibitem{Moreno GM KL 2003}
P.\ J.\ Moreno, P.\ P.\ Ho, and N.\ Vasconcelos, ``A
Kullback-Leibler divergence based kernel for SVM classification in
multimedia applications,'' {\em Advances in Neural Information
Processing Systems}, vol. 16, pp. 1385-1393, 2003.

\bibitem{Motley Keenan PL 1988}
A.\ J.\ Motley and J.\ P.\ Keenan, ``Personal communication radio
coverage in buildings at 900 MHz and 1700 MHz,'' {\em Electronics
Letters}, vol. 24, pp. 763-764, 1988.

\bibitem{Niu Chen Varshney T-SP 2006}
R.\ Niu, B.\ Chen, and P.\ K.\ Varshney, ``Fusion of decisions
transmitted over Rayleigh fading channels in wireless sensor
networks,'' {\em IEEE Trans. Signal Processing}, vol. 54, no. 3,
pp. 1018-1027, Mar. 2006.

\bibitem{Poor_book}
H.\ V.\ Poor, {\em An Introduction to Signal Detection and
Estimation}, Springer, New York, 1994.

\bibitem{Poor Thomas 1977}
H.\ V.\ Poor and J.\ B.\ Thomas, ``Applications of Ali-Silvey
distance measures in the design of generalized quantizers for
binary decision systems,'' {\em IEEE Trans. Commun.}, vol. 25, no.
9, pp. 893-900, Sept. 1977.

\bibitem{Predd Kulkarni Poor 2006}
J.\ B.\ Predd, S.\ R.\ Kulkarni, and H.\ V.\ Poor, ``Consistency
in models for distributed learning under communication
constraints,'' {\em IEEE Trans. Inform. Theory}, vol. 52, no. 1,
pp. 52-63, Jan. 2006.

\bibitem{Rago Willett Bar-Shalom 1996}
C.\ Rago, P.\ Willett, and Y.\ Bar-Shalom, ``Censoring sensors: A
low-communication-rate scheme for distributed detection,'' {\em
IEEE Trans. Aerosp. Electron. Syst.}, vol. 32, no. 2, pp. 554-568,
Apr. 1996.

\bibitem{Singer Warmuth GM KL 1998} Y.\ Singer and M.\ K.\ Warmuth,
``Batch and on-line parameter estimation of Gaussian mixtures
based on the joint entropy,'' {\em Advances in Neural Information
Processing Systems}, vol. 11, pp. 578-584, 1998.


\bibitem{Tse_Viswanath_book}
D.\ Tse and P.\ Viswanath {\em Fundamentals of Wireless
Communication}, Cambridge University Press: Cambeidge, UK, 2004.

\bibitem{Tsitsiklis 1993}
J.\ N.\ Tsitsiklis, ``Decentralized detection,'' {\em Advances in
Statistical Signal Proccessing, Signal Detection}, vol. 2, H.\ V.\
Poor and J.\ B.\ Thomas, Eds, pp. 297-344, JAI Press, Greenwich,
CT, 1993.

\bibitem{Varshney_book}
P.\ K.\ Varshney, {\em Distributed Detection and Data Fusion},
Springer, New York, 1996.

\bibitem{Viswanathan Varshney 1997}
R.\ Viswanathan and P.\ K.\ Varshney, ``Distributed detection with
multiple sensors: Part I -- Fundamentals,'' {\em Proc. IEEE}, vol.
85, no. 1, pp. 54-63, Jan. 1997.

\bibitem{Verdu_book}
S.\ Verd\'u, {\em Multiuser Detection}, Cambridge University
Press: Cambridge, UK, 1998.

\bibitem{Wang_Poor_book}
X.\ Wang and H.\ V.\ Poor, {\em Wireless Communication Systems:
Advanced Techniques for Signal Reception}, Prentice-Hall: Upper
Saddle River, NJ, 2004.


\end{thebibliography}
\end{document}